\documentclass[aps,prd,nofootinbib,showpacs,superscriptaddress, preprint]{revtex4-1}
\pdfoutput=1
\usepackage[utf8]{inputenc}
\usepackage{amsmath,amssymb}
\usepackage{graphicx}
\usepackage[usenames,dvipsnames]{color}
\usepackage{float}
\usepackage{bbold}
\usepackage[colorlinks,citecolor=blue]{hyperref}
\usepackage{color}
\usepackage{subfigure}

\begin{document}
\title{Observable Gravitational Waves in Minimal Scotogenic Model}
\author{Debasish Borah}
\email{dborah@iitg.ac.in}
\affiliation{Department of Physics, Indian Institute of Technology Guwahati, Assam 781039, India}

\author{Arnab Dasgupta}
\email{arnabdasgupta@protonmail.ch}
\affiliation{Institute of Convergence Fundamental Studies , Seoul-Tech, Seoul 139-743, Korea}
\author{Kohei Fujikura}
\email{fuji@th.phys.titech.ac.jp}
\affiliation{Department of Physics, Tokyo Institute of Technology, 2-12-1 Ookayama, Meguro-ku, Tokyo 152- 8551, Japan}
\author{Sin Kyu Kang}
\email{skkang@seoultech.ac.kr}
\affiliation{School of Liberal Arts, Seoul-Tech, Seoul 139-743, Korea}
\author{Devabrat Mahanta}
\email{devab176121007@iitg.ac.in}
\affiliation{Department of Physics, Indian Institute of Technology Guwahati, Assam 781039, India}

\begin{abstract}
We scrutinise the widely studied minimal scotogenic model of dark matter (DM) and radiative neutrino mass from the requirement of a strong first order electroweak phase transition (EWPT) and observable gravitational waves at future planned space based experiments. The scalar DM scenario is similar to inert scalar doublet extension of standard model where a strong first order EWPT favours a portion of the low mass regime of DM which is disfavoured by the latest direct detection bounds. In the fermion DM scenario, we get newer region of parameter space which favours strong first order EWPT as the restriction on mass ordering within inert scalar doublet gets relaxed. While such leptophilic fermion DM remains safe from stringent direct detection bounds, newly allowed low mass regime of charged scalar can leave tantalising signatures at colliders and can also induce charged lepton flavour violation within reach of future experiments. While we get such new region of parameter space satisfying DM relic, strong first order EWPT with detectable gravitational waves, light neutrino mass and other relevant constraints, we also improve upon previous analysis in similar model by incorporating appropriate resummation effects in effective finite temperature potential.
\end{abstract}

\maketitle

\section{Introduction}
\label{sec:intro}
The scotogenic framework, proposed by Ma in 2006  \cite{Ma:2006km} has been one of the most popular extensions of the standard model (SM) which simultaneously account for the origin of light neutrino mass and dark matter (DM). In spite of significant experimental evidences confirming the non-vanishing yet tiny neutrino masses and the presence of a mysterious, non-luminous, non-baryonic form of matter, dubbed as DM \cite{Tanabashi:2018oca}, its origin remains unaddressed in the SM. While the latest experimental constraints on light neutrino parameters can be obtained from recent global fits \cite{deSalas:2017kay, Esteban:2018azc}, the present DM abundance is quantified in  terms of density 
parameter $\Omega_{\rm DM}$ and $h = {\rm Hubble \; Parameter\; ({\bf H})}/(100 \;\text{km} ~\text{s}^{-1} 
\text{Mpc}^{-1})$ as 
\cite{Aghanim:2018eyx}:
$\Omega_{\text{DM}} h^2 = 0.120\pm 0.001$
at 68\% CL. The possibility of scalar DM in this model has been already studied in several works including \cite{Cirelli:2005uq, Barbieri:2006dq, Ma:2006fn, LopezHonorez:2006gr,  Hambye:2009pw, Dolle:2009fn, Honorez:2010re, LopezHonorez:2010tb, Gustafsson:2012aj, Goudelis:2013uca, Arhrib:2013ela, Dasgupta:2014hha, Diaz:2015pyv, Borah:2017dfn, Borah:2018rca}, whereas that of thermal or non-thermal fermion DM has also been studied \cite{Ahriche:2017iar, Mahanta:2019gfe}. The model can also account for the observed baryon asymmetry through successful leptogenesis (for a review of leptogenesis, see \cite{Davidson:2008bu}) in variety of different ways \cite{Hugle:2018qbw, Borah:2018rca,Huang:2018vcr, Baumholzer:2018sfb, Borah:2018uci, Mahanta:2019gfe, Mahanta:2019sfo}. While the observational evidences suggesting the presence of DM are purely based on its gravitational interactions, most of the particle DM models (including the scotogenic model) adopt a weak portal (but much stronger than gravitational coupling) between DM and the visible matter or the SM particles. If DM is of thermal nature, like in the weakly interacting massive particle (WIMP) paradigm (see \cite{Arcadi:2017kky} for a recent review), then such DM-SM couplings can be as large as the electroweak couplings and hence such DM can leave imprint on direct search experiments. However, none of the direct detection experiments like LUX \cite{Akerib:2016vxi}, PandaX-II \cite{Tan:2016zwf, Cui:2017nnn} and Xenon1T \cite{Aprile:2017iyp, Aprile:2018dbl} have reported any positive signal yet, giving more and more stringent upper bounds on DM-nucleon interactions.

Although the ongoing direct detection experiments have not exhausted all parameter space yet, specially in the context of the minimal scotogenic model, it is equally important to look for complementary probes. Scalar DM in this model is not only tightly constrained by direct detection experiments mentioned above, but can also be probed at indirect detection experiments looking for excess in gamma rays \cite{Borah:2017dfn} or X-rays \cite{Baumholzer:2019twf} depending upon the particular DM candidate chosen. Apart from usual collider signatures of the scalar doublet like dijet or dileptons plus missing energy \cite{Miao:2010rg, Gustafsson:2012aj, Datta:2016nfz, Poulose:2016lvz, Belyaev:2016lok, Belyaev:2018ext}, the scotogenic model can also give rise to exotic signatures like displaced vertex and disappearing or long-lived charged track for compressed mass spectrum of inert scalars and singlet fermion DM \cite{Borah:2018smz}. Apart from collider signatures of inert scalars in the model, as discussed in several earlier works mentioned above, the model can also have interesting signatures at lepton flavour violating decays like $\mu \rightarrow e \gamma, \mu \rightarrow e e e$ and $\mu \rightarrow e$ conversions \cite{Toma:2013zsa, Vicente:2014wga}. In this work, we examine one more way of probing this popular and predictive model indirectly. One such indirect possibility is to search for gravitational wave (GW) which has gained lots of attention due to planned near future experiments like LISA and other similar space based interferometer experiments~\cite{Team:1998rwe,Audley:2017drz,Belgacem:2019pkk,Armano:2019cac,Caprini:2015zlo,Caprini:2019egz, Kudoh:2005as, Sato:2017dkf}.\footnote{These new sensitivity curves are presented in Refs.~\cite{Alanne:2019bsm,Schmitz:2020syl}.}
A possible source of GW signals is a strong first-order phase transition (SFOPT) where, in particular, GW signals are generated by bubble collisions~\cite{Turner:1990rc,Kosowsky:1991ua,Kosowsky:1992rz,Kosowsky:1992vn,Turner:1992tz}, the sound wave of the plasma~\cite{Hindmarsh:2013xza,Giblin:2014qia,Hindmarsh:2015qta,Hindmarsh:2017gnf} and the turbulence of the plasma~\cite{Kamionkowski:1993fg,Kosowsky:2001xp,Caprini:2006jb,Gogoberidze:2007an,Caprini:2009yp,Niksa:2018ofa}. 
The study of  SFOPT is well-motivated in beyond the SM (BSM) with extended Higgs sectors, which allows for new source of CP violation and could produce baryon excess during  the electroweak phase transition (EWPT).
Therefore, it would be interesting to investigate if SFOPTcan happen at high temperature in BSM.
For a review of cosmological phase transitions and their experimental signatures, one may refer to the recent review article ~\cite{Mazumdar:2018dfl}.
While a strong first-order  EWPT  is not possible in the SM alone due to high mass of SM like Higgs as confirmed by lattice studies~\cite{Csikor:1998eu,Rummukainen:1998as}, addition of other scalars like the one in scotogenic model can allow the EWPT to be of first order. Several works have studied such interplay of DM and first order EWPT, specially in the presence of additional scalar doublet DM like we have in the present model \cite{Chowdhury:2011ga, Borah:2012pu, Gil:2012ya, Cline:2013bln, Blinov:2015vma, Huang:2017rzf, Liu:2017gfg}. While scalar doublet DM extension of the SM, popularly known as inert doublet model (IDM) allows a large parameter space supporting a SFOPT, most of this parameter space corresponds to sub-dominant DM \cite{Cline:2013bln} leaving a narrow window in low mass DM regime where both DM relic and SFOPT criteria can be simultaneously satisfied \cite{Borah:2012pu}. On the other hand, this low mass regime is getting increasingly in tension with direct search experiments as well as the collider constraints on invisible decay rate of the SM like Higgs, where the former remains much more stringent. In fact, we show that the parameter space of scalar DM which satisfies SFOPT is already disfavoured by Xenon1T data of 2018. Extending IDM to scotogenic model not only addresses the origin of light neutrino mass but also enlarges the parameter space that can simultaneously satisfy DM relic as well as SFOPT. We discuss the possibilities of both scalar and fermion DM in this model and constrain the parameter space from DM relic, SFOPT as well as light neutrino masses while incorporating relevant experimental and theoretical bounds. We then study the possibility of generating GW from such SFOPT and discuss the possibility of its detection in future experiments.\footnote{See e.g. Ref.~\cite{Hashino:2018wee} for the detailed analysis of constraints coming from GW detections.}

This paper is organised as follows. In section \ref{sec:model} we discuss the model, particle spectrum and different constraints. In section \ref{sec:PT} we review on the details of SFOPT and generation of GW. In section \ref{sec:DM} we briefly show how to calculate DM relic following thermal WIMP paradigm and then discuss our results in section \ref{sec:results}. We finally conclude in section \ref{sec:conclude}.


%
\section{Minimal Scotogenic Model}
\label{sec:model}
The minimal scotogenic model is an extension of the SM by three copies of SM-singlet fermions $N_i$ (with $i=1,2,3$) and one $SU(2)_L$-doublet scalar field $S$ ( called inert scalar doublet ), all being odd under an in-built and unbroken $Z_2$ symmetry, while the SM fields remain $Z_2$-even under the $Z_2$-symmetry such that
\begin{align}
N_i \rightarrow -N_i, \quad  S  \rightarrow -S, \quad \Phi \rightarrow \Phi, \quad \Psi_{\rm SM} \to 
\Psi_{\rm SM} \, ,
\label{eq:Z2}
\end{align}
where $\Phi$ is the SM Higgs doublet and $\Psi_{\rm SM}$'s stand for the SM fermions. The unbroken $Z_2$ symmetry not only prevents the inert scalar doublet to acquire any non-zero vacuum expectation value (VEV) but also forbids Yukawa coupling of lepton doublets with $Z_2$ odd singlet fermions via SM Higgs, keeping neutrinos massless at tree level. The Yukawa interactions of the lepton sector can be written as 
\begin{equation}\label{IRHYukawa}
{\cal L} \ \supset \ \frac{1}{2}(M_N)_{ij} N_iN_j + \left(Y_{ij} \, \bar{L}_i \tilde{S} N_j  + \text{h.c.} \right) \ . 
\end{equation}

The tree level scalar potential of the model, $V_{\rm tree}$, is given by
\begin{align}
V_{\rm tree} =& \lambda_{\rm SM} \left( \Phi^{\dag i} \Phi_i -\frac{v_{\rm SM}}{2}\right)^2 + m_1^2 S^{\dag i} S_i +\lambda_1 \left( \Phi^{\dag i} \Phi_i \right) \left( S^{\dag j} S_j\right) + \lambda_2 \left(\Phi^{\dag i} S_i\right) \left( S^{\dag j}\Phi_j\right) \nonumber \\
&+ \left[ \lambda_3 \Phi^{\dag i} \Phi^{\dag j} S_i S_j + {\rm h.c.}\right]+ \lambda_S \left(S^{\dag j}S_j\right)^2 . \label{eq:tree potential}
\end{align}
Here, indices $i$ and $j$ represent $SU(2)$ isospin. The $Z_2$ symmetry prevents linear and trilinear terms of the inert doublet with the SM Higgs. The bare mass squared term of the inert scalar doublet is chosen to be positive definite in order to ensure that it does not acquire any non-zero VEV. Absence of linear terms ensures that it does not even acquire any induced VEV after electroweak symmetry breaking (EWSB).
The first term in Eq.~\eqref{eq:tree potential} describes EWSB with $v_{\rm SM} \simeq 246$GeV and $\lambda_{\rm SM} \simeq 0.13$.
To compute mass spectrums of $H$ and $S$ fields, let us parametrise those fields as follows.
\begin{align}
\Phi= 
\begin{pmatrix}
0\\
\frac{\phi}{\sqrt 2}
\end{pmatrix}, \qquad S=
\begin{pmatrix}
S^+ \\
S_0
\end{pmatrix}.
\end{align}

For $S^+$ and $S_0$ fields, we have following mass terms:

\begin{align}
m_{\pm}^2 = m_1^2 +\frac{1}{2} \lambda_1 \phi^2,~
\mathcal{M}^2_{S_0} 
=
\begin{pmatrix}
m_1^2 +\dfrac{1}{2}\left(\lambda_1 +\lambda_2+2\lambda_3 \right) \phi^2,~0 \\
0,~m_1^2 +\dfrac{1}{2}\left(\lambda_1 +\lambda_2 -2\lambda_3 \right) \phi^2
\end{pmatrix}.
\end{align}
Using $S_0 = (H+iA)/\sqrt{2}$ we obtain the physical masses as
\begin{align}
&n_{\pm}= 2~:~m_\pm^2 =m_1^2+ \frac{1}{2} \lambda_1 \phi^2,\nonumber\\
&n_H=1~:~m_H^2 = m_1^2 +\frac{1}{2} \left(\lambda_1 +\lambda_2 +2\lambda_3 \right)\phi^2, \nonumber\\
&n_A=1~:~m_A^2 = m_1^2 +\frac{1}{2} \left(\lambda_1 +\lambda_2 -2\lambda_3\right)\phi^2, \label{eq:field dependent mass of inert doublet}
\end{align}
where $m_H~(m_A)$ and $m_{\pm}$ are the masses of CP-even (odd) component and the charged component of the inert scalar doublet, respectively, and $n_{\pm},~n_H$ and $n_A$ represent degrees of freedom (dof) of each fields. Present masses are obtained by $\phi = v_{\rm SM}$. In this notation, the SM Higgs mass is $m^2_h = 2 \lambda_{\rm SM} \phi^2$.

While neutrinos remain massless at tree level, a non-zero mass can be generated at one-loop level given by ~\cite{Ma:2006km, Merle:2015ica}
\begin{align}
(M_{\nu})_{ij} \ & = \ \sum_k \frac{Y_{ik}Y_{jk} M_{k}}{32 \pi^2} \left ( \frac{m^2_{H}}{m^2_{H}-M^2_k} \: \text{ln} \frac{m^2_{H}}{M^2_k}-\frac{m^2_{A}}{m^2_{A}-M^2_k}\: \text{ln} \frac{m^2_{A}}{M^2_k} \right) \nonumber \\ 
& \ \equiv  \ \sum_k \frac{Y_{ik}Y_{jk} M_{k}}{32 \pi^2} \left[L_k(m^2_{H})-L_k(m^2_{A})\right] \, ,
\label{numass1}
\end{align}
where 
$M_k$ is the mass eigenvalue of the mass eigenstate $N_k$ in the internal line and the indices $i, j = 1,2,3$ run over the three neutrino generations as well as three copies of $N_i$. The function $L_k(m^2)$ is defined as 
\begin{align}
L_k(m^2) \ = \ \frac{m^2}{m^2-M^2_k} \: \text{ln} \frac{m^2}{M^2_k} \, .
\label{eq:Lk}
\end{align}
It is important to ensure that the choice of Yukawa couplings as well as other parameters involved in light neutrino mass are consistent with the cosmological upper bound on the sum of neutrino masses, $\sum_i m_{i}\leq 0.11$ eV~\cite{Aghanim:2018eyx}, as well as the neutrino oscillation data~\cite{deSalas:2017kay, Esteban:2018azc}. In order to incorporate these constraints on model parameters, it is often useful to rewrite the neutrino mass formula given in equation \eqref{numass1} in a form resembling the type-I seesaw formula: 
\begin{align}
M_\nu \ = \ Y {\Lambda}^{-1} Y^T \, ,
\label{eq:nu2}
\end{align}
where we have introduced the diagonal matrix $\Lambda$ with elements
\begin{align}
 \Lambda_i \ & = \ \frac{2\pi^2}{\lambda_3}\zeta_i\frac{M_i}{v^2} \, , \\
\textrm {and}\quad \zeta_i & \ = \  \left(\frac{M_{i}^2}{8(m_{H}^2-m_{A}^2)}\left[L_i(m_{H}^2)-L_i(m_{A}^2) \right]\right)^{-1} \, . \label{eq:zeta}
\end{align}
The light neutrino mass matrix~\eqref{eq:nu2} which is complex symmetric, can be diagonalised by the usual Pontecorvo-Maki-Nakagawa-Sakata (PMNS) mixing matrix $U$ \footnote{Usually, the leptonic mixing matrix is given in terms of the charged lepton diagonalising matrix $(U_l)$ and light neutrino diagonalising matrix $U_{\nu}$ as $U = U^{\dagger}_l U_{\nu}$. In the simple case where the charged lepton mass matrix is diagonal which is true in our model, we can have $U_l = \mathbb{1}$. Therefore we can write $U = U_{\nu}$.}, written in terms of neutrino oscillation data (up to the Majorana phases) as
\begin{equation}
U=\left(\begin{array}{ccc}
c_{12}c_{13}& s_{12}c_{13}& s_{13}e^{-i\delta}\\
-s_{12}c_{23}-c_{12}s_{23}s_{13}e^{i\delta}& c_{12}c_{23}-s_{12}s_{23}s_{13}e^{i\delta} & s_{23}c_{13} \\
s_{12}s_{23}-c_{12}c_{23}s_{13}e^{i\delta} & -c_{12}s_{23}-s_{12}c_{23}s_{13}e^{i\delta}& c_{23}c_{13}
\end{array}\right) U_{\text{Maj}}
\label{PMNS}
\end{equation}
where $c_{ij} = \cos{\theta_{ij}}, \; s_{ij} = \sin{\theta_{ij}}$ and $\delta$ is the leptonic Dirac CP phase. The diagonal matrix $U_{\text{Maj}}=\text{diag}(1, e^{i\xi_1}, e^{i\xi_2})$ contains the undetermined Majorana CP phases $\xi_1, \xi_2$. The diagonal light neutrino mass matrix is therefore,
\begin{align}
D_\nu \ &= \ U^\dag M_\nu U^* \ = \ \textrm{diag}(m_1,m_2,m_3) \, .
\end{align}   
Since the inputs from neutrino data are only in terms of the mass squared differences and mixing angles, it would be  
useful for our purpose to express the Yukawa couplings in terms of light neutrino parameters. This is possible through the 
Casas-Ibarra (CI) parametrisation \cite{Casas:2001sr} extended to radiative seesaw model \cite{Toma:2013zsa} which 
allows us to write the Yukawa coupling matrix satisfying the neutrino data as
\begin{align}
Y \ = \ U D_\nu^{1/2} R^{\dagger} \Lambda^{1/2} \, ,
\label{eq:Yuk}
\end{align}
where $R$ is an arbitrary complex orthogonal matrix satisfying $RR^{T}=\mathbb{1}$.


\subsection{Constraints on Model Parameters}
\label{sec:constraint}
Precision measurements at LEP experiment forbids additional decay channels of the SM gauge bosons. For example, it strongly constrains the decay channel $Z \rightarrow H A$ requiring $m_H + m_A > m_Z$. Additionally, LEP precision data also rule out the region $m_H < 80 \; {\rm GeV}, m_A < 100 \; {\rm GeV}, m_A - m_H > 8 {\rm GeV}$ \cite{Lundstrom:2008ai}. We take the lower bound on charged scalar mass $m_{\pm} > 90$ GeV. If $m_{H, A} < m_h/2$, the large hadron collider (LHC) bound on invisible Higgs decay comes into play \cite{Aaboud:2019rtt} which can constrain the SM Higgs coupling with H, A namely $\lambda_1 +\lambda_2 \pm 2\lambda_3$ to as small as $10^{-4}$, which however remains weaker than DM direct detection bounds in this mass regime (see for example, \cite{Borah:2017hgt}).

The LHC experiment can also put bounds on the scalar masses in the model, though in a model dependent way. Depending upon the mass spectrum of its components, the heavier ones can decay into the lighter ones and a gauge boson, which finally decays into a pair of leptons or quarks. Therefore, we can have either pure leptonic final states plus missing transverse energy (MET), hadronic final states plus MET or a mixture of both. The MET corresponds to DM or light neutrinos. In several earlier works \cite{Miao:2010rg, Gustafsson:2012aj, Datta:2016nfz}, the possibility of opposite sign dileptons plus MET was discussed. In \cite{Poulose:2016lvz}, the possibility of dijet plus MET was investigated with the finding that inert scalar masses up to 400 GeV can be probed at high luminosity LHC. In another work \cite{Hashemi:2016wup}, tri-lepton plus MET final states was also discussed whereas mono-jet signatures have been studied by the authors of \cite{Belyaev:2016lok, Belyaev:2018ext}. The enhancement in dilepton plus MET signal in the presence of additional vector like singlet charged leptons was also discussed in \cite{Borah:2017dqx}. Exotic signatures like displaced vertex and disappearing or long-lived charged track for compressed mass spectrum of inert scalars and singlet fermion DM was studied recently by the authors of \cite{Borah:2018smz}. 

%
%

In addition to the collider or direct search constraints, there exists theoretical constraints also. For instance, the scalar potential of the model should be bounded from below in any field direction. This criteria leads to the following co-positivity conditions.~\cite{Sher:1988mj,Branco:2011iw,Goudelis:2013uca,Dercks:2018wch}:
\begin{align}
\lambda_S > 0,~\lambda_1+2\sqrt{\lambda_{\rm SM} \lambda_S} >0,~\lambda_1+\lambda_2 -\frac{|\lambda_3|}{2}+2\sqrt{\lambda_{\rm SM}\lambda_S} > 0.
\end{align}
The last condition comes from unitarity of the $S$-matrix elements~\cite{Ginzburg:2005dt,Aoki:2012yt}.
The coupling constants appeared in above expressions are evaluated at the electroweak scale, $v_{\rm SM}$. Also, in order to avoid perturbative breakdown, all dimensionless couplings like quartic couplings $(\lambda_i)$, Yukawa couplings $(Y_{ij})$, gauge couplings $(g_i)$  should obey the the perturbativity conditions:
\begin{align}
 |\lambda_i| < 4\pi, \lvert Y_{ij} \rvert < \sqrt{4\pi}, g_i < \sqrt{4\pi}
\end{align}
where indices run over appropriate numbers according to the Lagrangian.

\section{SFOPT and production of GW}
\label{sec:PT}

The main purpose of this section is to show how to calculate the finite-temperature effective potential and GW signals generated by a first-order phase transition.
The crucial difference from the previous analysis done in Ref.~\cite{Borah:2012pu} is that we here include effects of resummation in order to account for IR divergences in finite-temperature field theory, and explicitly calculate GW signals.

\subsection{Finite-temperature effective potential}\label{sec:FOPT}

In this subsection, we calculate the one-loop effective potential at finite-temperature.
A total effective potential can be schematically divided into following form:
\begin{align}
V_{\rm tot} = V_{\rm tree} + V_{\rm CW} +V_{\rm th},
\end{align}
where $V_{\rm tree},~V_{\rm CW}$ and $V_{\rm th}$ denote the tree level scalar potential, the one-loop Coleman-Weinberg potential, the thermal effective potential, respectively.
The tree level scalar potential is given by Eq.~\eqref{eq:tree potential}.
In finite-temperature field theory, the effective potential, $V_{\rm CW}$ and $V_{\rm thermal}$, are calculated by using standard background field method~\cite{Dolan:1973qd,Quiros:1999jp}.
In the following calculations, we take Landau gauge for simplicity.\footnote{The gauge dependence of the thermal effective potential is discussed by many authors. See. e.g. Refs.~\cite{Wainwright:2011qy,Wainwright:2012zn} and references therein.}

The Coleman-Weinberg potential~\cite{Coleman:1973jx} with $\overline{\rm DR}$ regularisation is given by
\begin{align}
V_{\rm CW} = \sum_i (-)^{n_{f}} \frac{n_i}{64\pi^2} m_i^4 (\phi) \left(\log\left(\frac{m_i^2 (\phi)}{\mu^2} \right)-\frac{3}{2} \right),
\end{align}
where suffix $i$ represents particle species, and $n_i,~m_i (\phi)$ are the degrees of freedom (d.o.f) and field dependent masses of $i$'th particle.
In addition, $\mu$ is the renormalisation scale, and $(-)^{n_f}$ is $+1$ for bosons and $-1$ for fermions, respectively.
In our analysis, for simplicity, we take the renormalisation scale as $\mu = v_{\rm SM}$ because the electroweak scale is the only relevant energy scale.
We confirm that nucleation temperatures are around at electroweak scale for the whole region of the parameter space, and thus, the one-loop calculation is validated with this choice.
(For the definition of the nucleation temperature, see Eq.~\eqref{eq:nucleation temperature} in the next subsection.)

Thermal contributions to the effective potential are given by
\begin{align}
V_{\rm th} = \sum_i \left(\frac{n_{\rm B_i}}{2\pi^2}T^4 J_B \left[\frac{m_{\rm B_i}}{T}\right] - \frac{n_{\rm F_{i}}}{2\pi^2}J_F \left[\frac{m_{\rm F_{i}}}{T}\right]\right),
\end{align}
where $n_{B_i}$ and $n_{F_i}$ denote the dof of the bosonic and fermionic particles, respectively.
In this expressions, $J_B$ and $J_F$ functions are defined by following functions:
\begin{align}
&J_B(x) =\int^\infty_0 dz z^2 \log\left[1-e^{-\sqrt{z^2+x^2}}\right] \label{eq:J_B},\\
&J_F(x) =   \int^\infty_0 dz z^2 \log\left[1+e^{-\sqrt{z^2+x^2}}\right].
\end{align}
In our analysis, we do not use high-temperature expansions only validated for small $x$ to evaluate $J_B (x)$ and $J_F(x)$.
One should note that a potential barrier arises from $J_B (x)$ which triggers a first-order phase transition.
Therefore, the strong first-order electroweak phase transition requires bosons strongly coupled to the Higgs field. (See next subsection for the definition of the SFOPT.)
In comparison to the SM, this model contains an additional $SU(2)_W$ inert scalar doublet, and thus, the strong first-order electroweak phase transition takes place as we will see Sec.~\ref{sec:results}. 

For a calculation of $V_{\rm th}$, we include a contribution from daisy diagram to improve the perturbative expansion during the phase transition~\cite{Fendley:1987ef,Parwani:1991gq,Arnold:1992rz}.
The daisy improved effective potential can be calculated by inserting thermal masses into the zero-temperature field dependent masses.
The author of Ref.~\cite{Parwani:1991gq} proposed the thermal resummation prescription in which the thermal corrected field dependent masses are used for the calculation in $V_{\rm CW}$ and $V_{\rm th}$ (Parwani method).
In comparison to this prescription, authors of Ref.~\cite{Arnold:1992rz} proposed alternative prescription for the thermal resummation (Arnold-Espinosa method).
They include the effect of daisy diagram only for Matsubara zero-modes inside $J_B$ function defined in Eq.~\eqref{eq:J_B}.
A qualitative difference between two prescriptions is discussed in Ref.~\cite{Basler:2016obg}.
For simplicity, we here use former (Parwani method) prescription in order to implement daisy resummation in the public code \texttt{CosmoTransitions}~\cite{Wainwright:2011kj}.

The thermal mass for the inert scalar doublet, $\Pi_S (T)$, is calculated in Ref.~\cite{Blinov:2015vma} as 
\begin{align}
\Pi_S = \left( \frac{1}{8}g_2^2+\frac{1}{16}(g_1^2 +g_2^2 ) +\frac{1}{2}\lambda_S + \frac{1}{12}\lambda_1 +\frac{1}{24}\lambda_A +\frac{1}{24}\lambda_{H}\right)T^2. \label{eq:thermal mass of eta}
\end{align}
where $\lambda_A = \lambda_1+\lambda_2 - 2\lambda_3$ and $\lambda_{H} = \lambda_1 +\lambda_2 +2\lambda_3$, 
and $g_1$ and $g_2$ are the gauge coupling of $SU(2)_W$ and $U(1)_Y$, respectively. One should note that the Yukawa coupling between the inert doublet and the neutrino given by Eq.~\eqref{IRHYukawa} may give an additional contribution to $\Pi_S (T)$. However, since the right-handed neutrino has Majorana mass comparable to the nucleation temperature, this effect would be suppressed due to the Boltzmann suppression in finite-temperature, and hence, we expect that this contribution will not change our result significantly. We therefore neglect this contribution in our analysis.

For gauge bosons, the longitudinal mode and the transverse mode obtain different thermal masses called Debye mass and the magnetic mass, respectively~\cite{Linde:1980ts}.
Since the magnetic mass is subdominant, we only calculate the Debye mass for $SU(2)_W\times U(1)_Y$ gauge bosons.
The Debye mass can be estimated by using SM result~\cite{Carrington:1991hz,Kapusta:2006pm} with adding an additional $SU(2)_W$ inert scalar doublet as
\begin{align}
\Pi_W = 2g_2^2 T^2,~\Pi_Y =2 g_1^2 T^2.
\end{align} 
The field dependent masses are obtained by diagonalising mass matrices for $SU(2)_W \times U(1)_Y$ gauge bosons and neutral component of the inert scalar doublet.

We consider contributions coming from inert scalar doublet, $SU(2)_W\times U(1)_Y$ gauge bosons and top quarks because those couplings are large compared to other particles.
Field dependent masses and d.o.f are then listed as
\begin{align}
 &n_{\pm} =2 : m_{\pm}^2 (\phi) = m_1^2+\frac{\lambda_1}{2} \phi^2 + \Pi_S (T),\\
 &n_{H}= 1:~ m_{H}^2 (\phi) = m_1^2 +\frac{1}{2}\lambda_{H}\phi^2 + \Pi_S (T),\\
 &n_A=1:~m_A^2 (\phi) = m_1^2 + \frac{1}{2}\lambda_A \phi^2 +\Pi_S (T),\\
 &n_W=4:~ m^2_W (\phi) = \frac{g_2^2}{4} \phi^2,\\
 &n_Z=2:~m_Z^2 (\phi)=\frac{(g_1^2+g_2^2)}{4}\phi^2,\\
 &n_{W_L}=2:~m^2_{W_L}(\phi) =\frac{g_2^2}{4}\phi^2 + \Pi_W (T),\\
 &n_{Z_L}=1:~m^2_{Z_L}(\phi) = \frac{1}{2} (m_Z^2 (\phi)+\Pi_W (T)+\Pi_Y (T)+\Delta(\phi,T) ) ,\\
 &n_{\gamma_L}=1:~m^2_{\gamma_L}(\phi) =  \frac{1}{2} (m_Z^2 (\phi)+\Pi_W (T)+\Pi_Y (T)-\Delta(\phi,T) ),\\
 &\Delta^2 (\phi,T)\equiv \left(\frac{g_2^2}{4}\phi^2+\Pi_W (T) -\frac{g_1^2}{4}\phi^2 -\Pi_Y(T)\right)^2 +4g_1^2 g_1^2 \phi^4.
\end{align}
The last contribution comes from the longitudinal mode of the massive photon.

\subsection{First-order phase transitions and GW signals}

In this subsection, we briefly review basics of the first-order phase transition and GW production.
When the phase transition proceeds via the tunnelling, we call it the first-order phase transition.
If the phase transition is of first order, the spherical symmetric field configurations called bubbles are nucleated in the early universe. Then they expand and eventually coalesce with each other.

The tunnelling rate per unit time and per unit volume, $\Gamma$, is given by
\begin{align}
\Gamma (T) = \mathcal{A}(T) e^{-\mathcal{B}},
\end{align}
where $\mathcal{A}(T)\sim T^4$ and $\mathcal{B}$ are determined by the dimensional analysis and given by the classical configurations, called bounce, respectively.
At finite-temperature, the $O(3)$ symmetric bounce solution is favoured~\cite{Linde:1980tt} for the case without significant supercooling.
The $O(3)$ symmetric bounce solution is calculated by solving the differential equation given by
\begin{align}
    \frac{d^2 \phi}{dr^2}+\frac{2}{r}\frac{d\phi}{dr} = \frac{\partial V_{\rm tot}}{\partial \phi}\label{eq:bounce diff},
\end{align}
with the following boundary conditions:
\begin{align}
\phi(r\to \infty)= \phi_{\rm false},~\left.\frac{d\phi}{dr}\right|_{r=0} =0.\label{eq:boundary condition}
\end{align}
In this expression, $\phi_{\rm false}$ denotes the position of the false vacuum.
Then the $O(3)$ symmetric bounce action is given by
\begin{align}
    S_3 =\int_0^{\infty} dr 4\pi r^2 \left[\frac{1}{2}\left(\frac{d\phi}{dr}\right)^2 +V_{\rm tot}(\phi,T)\right],
\end{align}
where $\phi$ satisfies equation of motion given by Eqs.~\eqref{eq:bounce diff} and \eqref{eq:boundary condition}.

The nucleation temperature defined as the temperature when bubbles are nucleated, $T_n$, is conservatively estimated by comparing the tunnelling rate to the Hubble parameter as
\begin{align}
    \Gamma (T_n) = {\bf H}^4(T_n).
\end{align}
Here, the Hubble parameter is given by ${\bf H}(T)\simeq 1.66\sqrt{g_*}T^2/M_{\rm Pl}$ with $g_*$ being the dof of the radiation component.
Then above condition turns out to be 
\begin{align}
    \frac{S_3(T_n)}{T_n} \simeq 140, \label{eq:nucleation temperature}
\end{align}
for $g_*\sim 100$ and $T_n \sim 100$ GeV.
We evaluate bounce action by using public code \texttt{cosmoTransitions}~\cite{Wainwright:2011kj} written in Python.
If $\phi(T_n) / T_n>1$ is satisfied, where $\phi(T_n)$ is the Higgs VEV at the nucleation temperature, $T=T_n$, the electroweak phase transition is called "strong" first order.

Amplitudes and frequencies of the GW are mainly determined by two parameters.
One is the ratio of the amount of vacuum energy released by the phase transition to the radiation energy density of the universe, $\rho_{\rm rad}= g_*\pi^2 T^4/30 $, given by
\begin{align}
    \alpha =\frac{\epsilon}{\rho_{\rm rad}},
\end{align}
with
\begin{align}
    \epsilon = \Delta V_{\rm tot} - \frac{T}{4} \frac{\partial \Delta V_{\rm tot}}{\partial T},
\end{align}
where $\Delta V_{\rm tot} \equiv V_{\rm tot}(\phi_{\rm false})- V_{\rm tot}(\phi_{\rm true})$ is the free energy difference between the false and true vacuum.
$\epsilon$ is related to the change in the trace of the energy-momentum tensor across the bubble wall (See e.g. Refs.~\cite{Caprini:2019egz,Giese:2020rtr} for detailed studies).

The another important parameter is duration of the phase transition denoted by $\beta$ which is the characteristic time scale of the phase transition.
$\beta$ is defined as the time variation of the tunnelling rate of the phase transition given by~\cite{Caprini:2015zlo}
\begin{align}
\frac{\beta}{{\bf H}(T)} \simeq T\frac{d}{dT} \left(\frac{S_3}{T} \right).
\end{align}
Both parameters $\alpha$ and $\beta$ are evaluated at $T=T_n$.

Let us next review the basic of GW signals generated by first-order phase transitions.
There are three sources producing GWs : bubble collisions, sound wave of the plasma, and turbulence of the plasma.
As confirmed in Refs.~\cite{Bodeker:2017cim,Ellis:2019oqb}, a contribution from bubble collisions is always subdominant for the case with small supercooling, $\alpha < 1$, since the bubble kinetic energy is almost converted into the thermal plasma due to the process called "transition radiation".
We therefore consider contributions coming from thermal plasma, especially, the sound wave and the turbulence.
Thus, total contribution to the GW signals can be decomposed into the following form:
\begin{align}
    \Omega_{\rm tot}h^2 \simeq \Omega_{\rm sound}h^2+\Omega_{\rm tur}h^2,
\end{align}
where $\Omega_{\rm sound}$ and $\Omega_{\rm tur}$ are the contributions from the sound wave and the turbulence of the plasma, respectively.

The contribution coming from sound wave of the plasma is estimated in Refs.~\cite{Caprini:2009yp,Binetruy:2012ze,Hindmarsh:2015qta,Caprini:2015zlo} as
\begin{align}
    \Omega_{\rm sound} h^2 = 2.65\times 10^{-6} \times {\bf H}\tau_{\rm sound} \left(\frac{{\bf H}(T_n)}{\beta}\right)\left(\frac{\kappa_{\rm sound} \alpha}{1+\alpha}\right)^2 \left(\frac{100}{g_*}\right)^{\frac{1}{3}} v_w \left(\frac{f}{f_{\rm sound}}\right)^3 \left[\frac{7}{4+3\left(\frac{f}{f_{\rm sound}}\right)^2} \right]^{\frac{7}{2}},
\end{align}
where $\kappa_{\rm sound}$ and $v_w$ are the efficiency factor and the bubble wall velocity, respectively.
The factor ${\bf H} \tau_{\rm sound}$ represents a suppression factor coming from the short-lasting sound wave as originally pointed out in Ref.~\cite{Ellis:2018mja} (Also, See e.g. Refs.~\cite{Cutting:2019zws,Ellis:2020awk}.) and is given by
\begin{align}
{\bf H} \tau_{\rm sound} = {\rm min} \left\{1,~ (8\pi)^{\frac{1}{3}} \left(\frac{{\rm max} \{ c_s,~v_w \}}{\beta / {\bf H}(T_n)}\right) \left(\frac{4}{3} \frac{1+\alpha}{\kappa_{\rm sound}\alpha}\right)^{\frac{1}{2}} \right\}.
\end{align}
Here $c_s$ is the speed of sound wave in the plasma and $f_{\rm sound}$ is the peak frequency of the GW signals given by
\begin{align}
    f_{\rm sound}=  1.9 \times 10^{-2}{\rm mHz} \times \frac{1}{v_w}\left( \frac{\beta}{{\bf H}(T_n)}\right) \left(\frac{T_n}{100{\rm GeV}}\right)\left(\frac{g_*}{100}\right)^{\frac{1}{6}}  .
\end{align}
We here simply estimate the bubble wall velocity by adopting following formula~\cite{Steinhardt:1981ct}\footnote{See Refs.~\cite{Huber:2013kj,Leitao:2014pda,Dorsch:2018pat,Cline:2020jre}, however, for the discussion of the bubble wall velocity $v_w$.}:
\begin{align}
    v_w = \frac{1/\sqrt{3} + \sqrt{\alpha^2 + 2\alpha/3}}{1+\alpha}.
\end{align}
With above bubble wall velocity called Jouguet detonations, the efficiency factor, $\kappa_{\rm sound}$, is fitted by following formula as found in Ref.~\cite{Espinosa:2010hh}:
\begin{align}
    \kappa_{\rm sound} = \frac{\sqrt{\alpha}}{0.135+\sqrt{0.98+\alpha}}   .
\end{align}

The contribution from turbulence plasma is estimated in Refs.~\cite{Caprini:2009yp,Binetruy:2012ze} as
\begin{align}
    \Omega_{\rm tur}h^2 = 3.35\times 10^{-4} \left( \frac{{\bf H}(T_n)}{\beta}\right) \left(\frac{\kappa_{\rm tur}\alpha}{1+\alpha} \right)^{\frac{3}{2}} \left(\frac{100}{g_*} \right)^{\frac{1}{3}} v_w \frac{\left(\frac{f}{f_{\rm tur}}\right)^3}{\left(1+ \left(\frac{f}{f_{\rm tur}} \right)\right)^{\frac{11}{3}}\left( 1+\frac{8\pi f}{H_0}\right)},
\end{align}
where $\kappa_{\rm tur}$ is the efficiency factor of the turbulence.
$\kappa_{\rm tur},~H_0$ and the peak frequency, $f_{\rm tur}$, are given by 
\begin{align}
&\kappa_{\rm tur} \simeq 0.1 \kappa_{\rm sound},\\
&H_0 \simeq 1.65\times 10^{-4}{\rm mHz} \times \left(\frac{T_n}{100{\rm GeV}} \right) \left(\frac{g_*}{100} \right)^{\frac{1}{6}} ,\\
&f_{\rm tur} =2.7\times 10^{-2} {\rm mHz} \left( \frac{1}{v_w}\right) \left( \frac{T_n}{100 {\rm GeV}} \right) \left(\frac{\beta}{{\bf H}(T_n)}\right) \left(\frac{g_*}{100} \right)^{\frac{1}{6}}.
\end{align}

\section{Dark Matter}
\label{sec:DM}
The lightest $Z_2$ odd particle, if electromagnetically neutral, is the DM candidate in the model. Among the two scalar DM candidates $H, A$, we consider $H$ as the DM, without any loss of generality. In the fermion DM scenario, the lightest right handed neutrino $N_1$ is the DM candidate. The relic abundance calculation of DM follows the usual approach of Boltzmann equation. If DM is of WIMP type and hence was produced thermally in the early universe, the evolution of its number density $n_{\rm DM}$ can be tracked by solving the Boltzmann equation
\begin{equation}
\frac{dn_{\rm DM}}{dt}+3 {\bf H}n_{\rm DM} \ = \ -\langle \sigma v \rangle \left[n^2_{\rm DM} -(n^{\rm eq}_{\rm DM})^2\right],
\label{eq:BE}
\end{equation}
where $n^{\rm eq}_{\rm DM}$ is the equilibrium number density of DM and $ \langle \sigma v \rangle $ is the thermally averaged annihilation cross section, given by~\cite{Gondolo:1990dk}
\begin{equation}
\langle \sigma v \rangle \ = \ \frac{1}{8m_{\rm DM}^4T K^2_2\left(\frac{m_{\rm DM}}{T}\right)} \int\limits^{\infty}_{4m_{\rm DM}^2}\sigma (s-4m_{\rm DM}^2)\sqrt{s}\: K_1\left(\frac{\sqrt{s}}{T}\right) ds \, ,
\label{eq:sigmav}
\end{equation}
where $K_i(x)$'s are modified Bessel functions of order $i$. In the presence of coannihilations, the effective cross section at freeze-out can be expressed as~\cite{Griest:1990kh}
\begin{align}
\sigma_{\rm eff} 
& \ = \ \sum_{i,j}^{N}\langle \sigma_{ij}v\rangle \frac{g_ig_j}{g^2_{\rm eff}}(1+\Delta_i)^{3/2}(1+\Delta_j)^{3/2}e^{-z_f(\Delta_i + \Delta_j)} \, , 
\end{align}
where $\Delta_i = \frac{m_i-m_{\text{DM}}}{m_{\text{DM}}}$ is the relative mass difference between the heavier component $i$ of the inert Higgs doublet (with $g_i$ internal dof) and the DM,  
\begin{align}
g_{\rm eff} & \ = \ \sum_{i=1}^{N}g_i(1+\Delta_i)^{3/2}e^{-z_f\Delta_i} 
\end{align}
is the total effective dof, and 
\begin{align}
\langle \sigma_{ij} v \rangle & \ = \ \frac{z_f}{8m^2_im^2_jm_{\text{DM}}K_2\left(\frac{m_iz_f}{m_{\text{DM}}}\right)K_2\left(\frac{m_j z_f}{m_{\text{DM}}}\right)} \nonumber \\
& \qquad \times \int\limits^{\infty}_{(m_i+m_j)^2}ds \: \sigma_{ij}\left(s-2(m_i^2+m_j^2)\right) \sqrt{s}\: K_1\left(\frac{\sqrt{s}\: z_f}{m_{\text{DM}}}\right) 
\label{eq:thcs}
\end{align}
is the modified thermally averaged cross section, compared to equation~\eqref{eq:sigmav}. In the above expressions
\begin{equation}
z_f \equiv \frac{m_{\rm DM}}{T_f} \ = \ \ln \left(0.038\frac{g}{\sqrt{g_*}}M_{\text{Pl}}m_{\rm DM}\langle \sigma v\rangle_f\right) \, , 
\label{xf}
\end{equation} 
with $g$ being the number of internal DOF of the DM and the subscript $f$ on $\langle \sigma v \rangle$ means that the quantity is evaluated at the freeze-out temperature $T_f$. The freeze-out temperature corresponds to the epoch where the rate of interaction equals that of the expansion, $ \Gamma (T=T_f) = {\bf H} (T=T_f)$. Since $N_1$ is a gauge singlet, there exists the possibility of non-thermal fermion DM also in this model, as discussed by \cite{Mahanta:2019gfe}. While the scalar DM can not be a purely non-thermal DM due to its gauge interactions, it can receive a non-thermal contribution from right handed neutrino decay at late epochs, as discussed by \cite{Borah:2017dfn}. We do not discuss such possibilities in this work, as it is unlikely to give new insights into the correlation between DM parameter space and that of SFOPT. For solving the Boltzmann equations relevant to DM, we have used \texttt{micrOMEGAs} package \cite{Belanger:2013oya}.

\section{Results and Discussion}
\label{sec:results}

In this section, we discuss our results obtained after performing a full numerical scan by incorporating all existing constraints and the criteria for a SFOPT and the correct DM relic density. In this parameter search, we vary the parameters $m_1,~\lambda_1,~\lambda_2,~\lambda_3$ and $\lambda_S$. While all four parameters are relevant for SFOPT, the last one does not affect the DM relic abundance. We have also imposed the LEP bounds as well as the perturbative and vacuum stability conditions discussed earlier. The constraints coming from light neutrino masses are incorporated by using Casas-Ibarra parametrisation discussed before.

We here qualitatively comment on the strength of the electroweak phase transition.
As we already noted in Sec.~\ref{sec:FOPT}, the strong first-order electroweak phase transition is realised when additional bosons present in the model strongly couple to the SM Higgs.
In this model, the inert scalar doublet, $S$, couples to the SM Higgs, and thus, larger $\lambda_H,~\lambda_A$ and $\lambda_1$ make the phase transition strength stronger.
Also, larger $m_1$ makes the phase transition strength weaker due to the screening effect~\cite{Quiros:1999jp}. For the same reason, since a larger $\lambda_S$ corresponds to larger thermal mass of the inert scalar (See Eq.~\eqref{eq:thermal mass of eta}), it makes the phase transition strength weaker. 
Therefore the strong first-order electroweak phase transition requires larger $\lambda_H,~\lambda_A$ and $\lambda_1$, and smaller $m_1$ and $\lambda_S$. Quantitative discussions are presented in each subsections below which are created according to the choice of DM candidate in the minimal scotogenic model.

\subsection{Scalar dark matter}

In this subsection, we show some results in the case of scalar DM scenario.
In addition to collider bounds, the perturbative and the vacuum stability conditions, we impose conditions $\lambda_3 <0$ and $\lambda_2 + 2\lambda_3 <0$ in order to make CP even component of inert doublet $H$ to be DM candidate. 

\begin{figure}[tbp]
\includegraphics[width=0.47\textwidth]{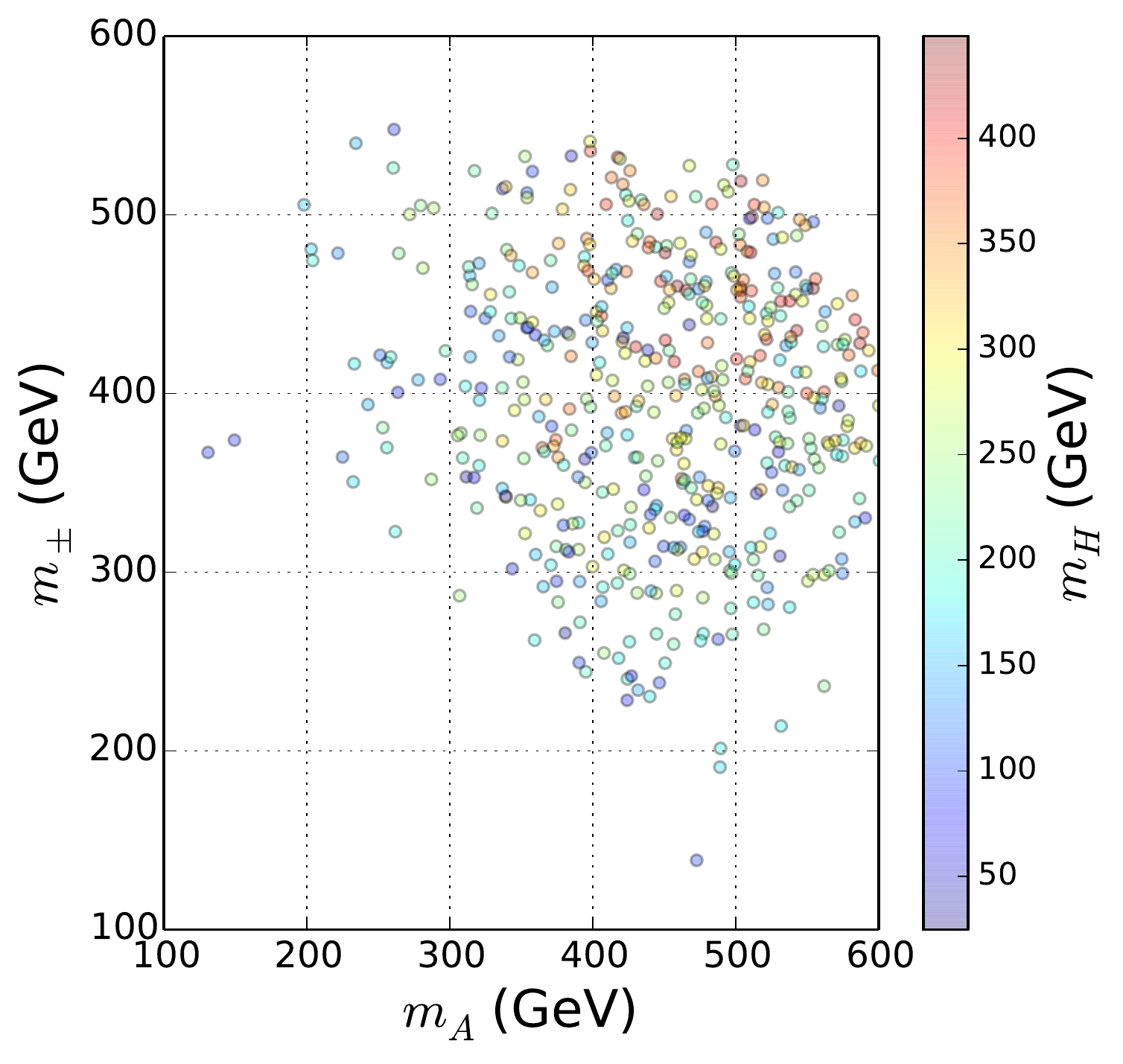}
\includegraphics[width=0.47\textwidth]{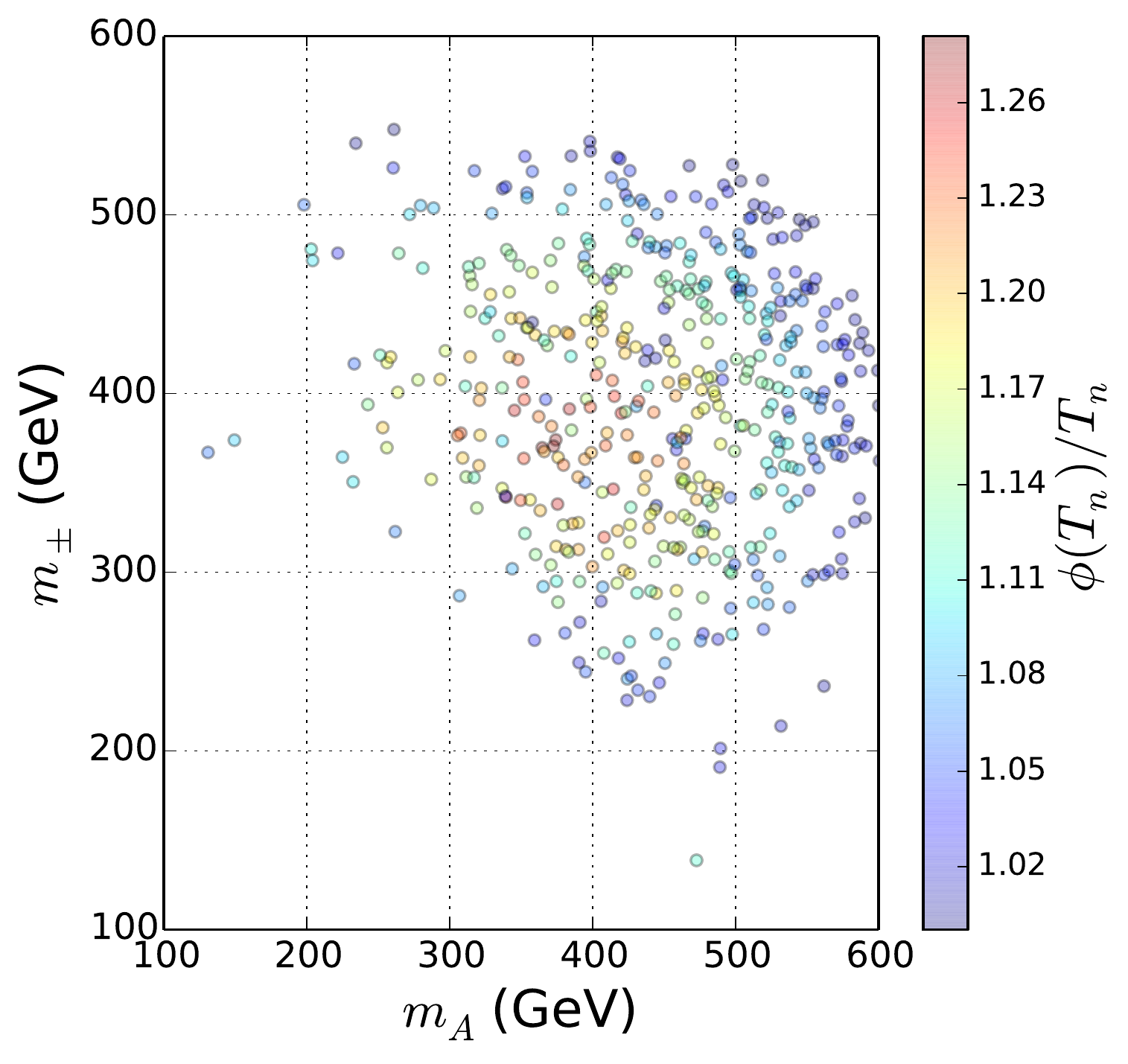}
\includegraphics[width=0.47\textwidth]{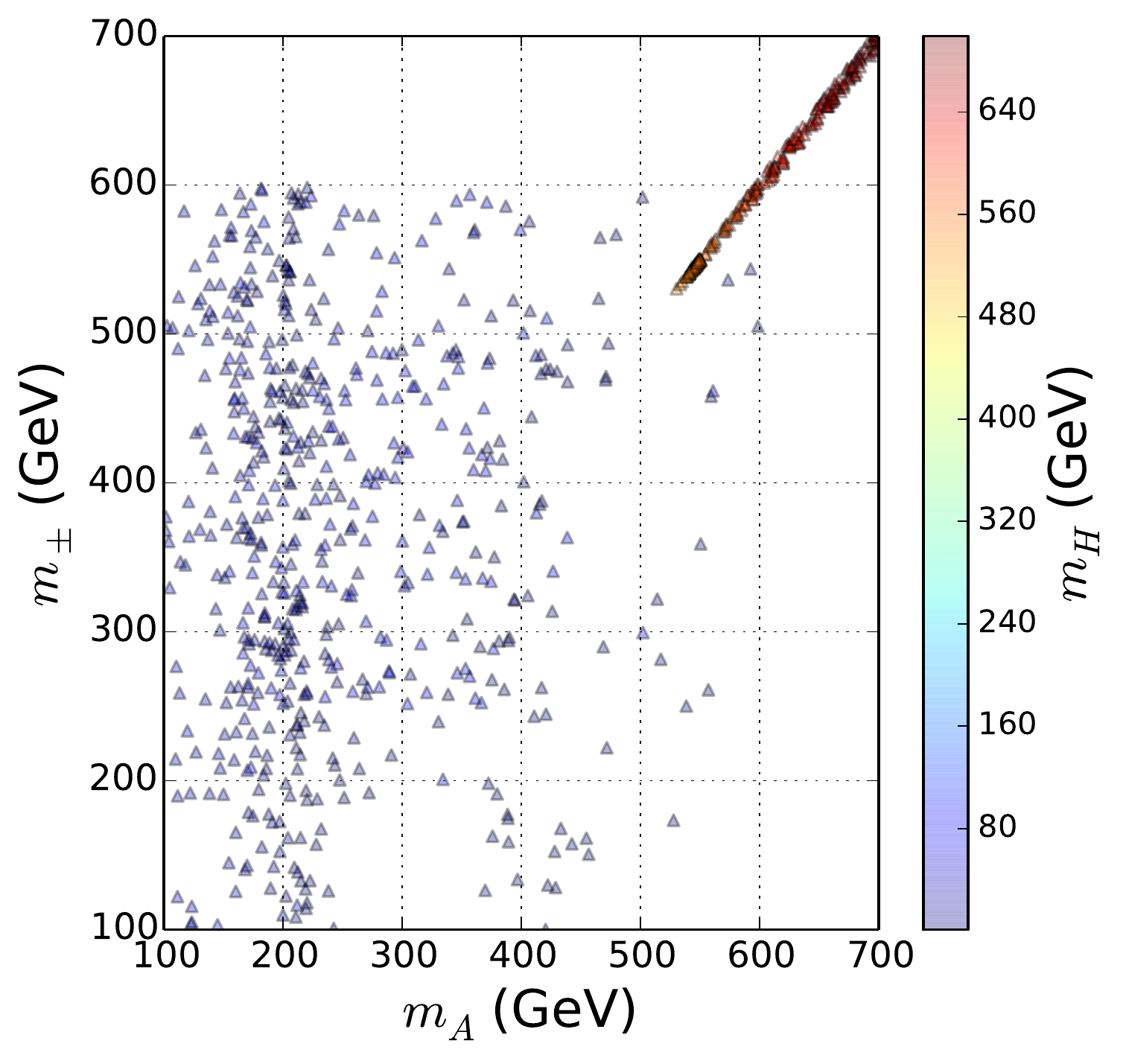}
\includegraphics[width=0.47\textwidth]{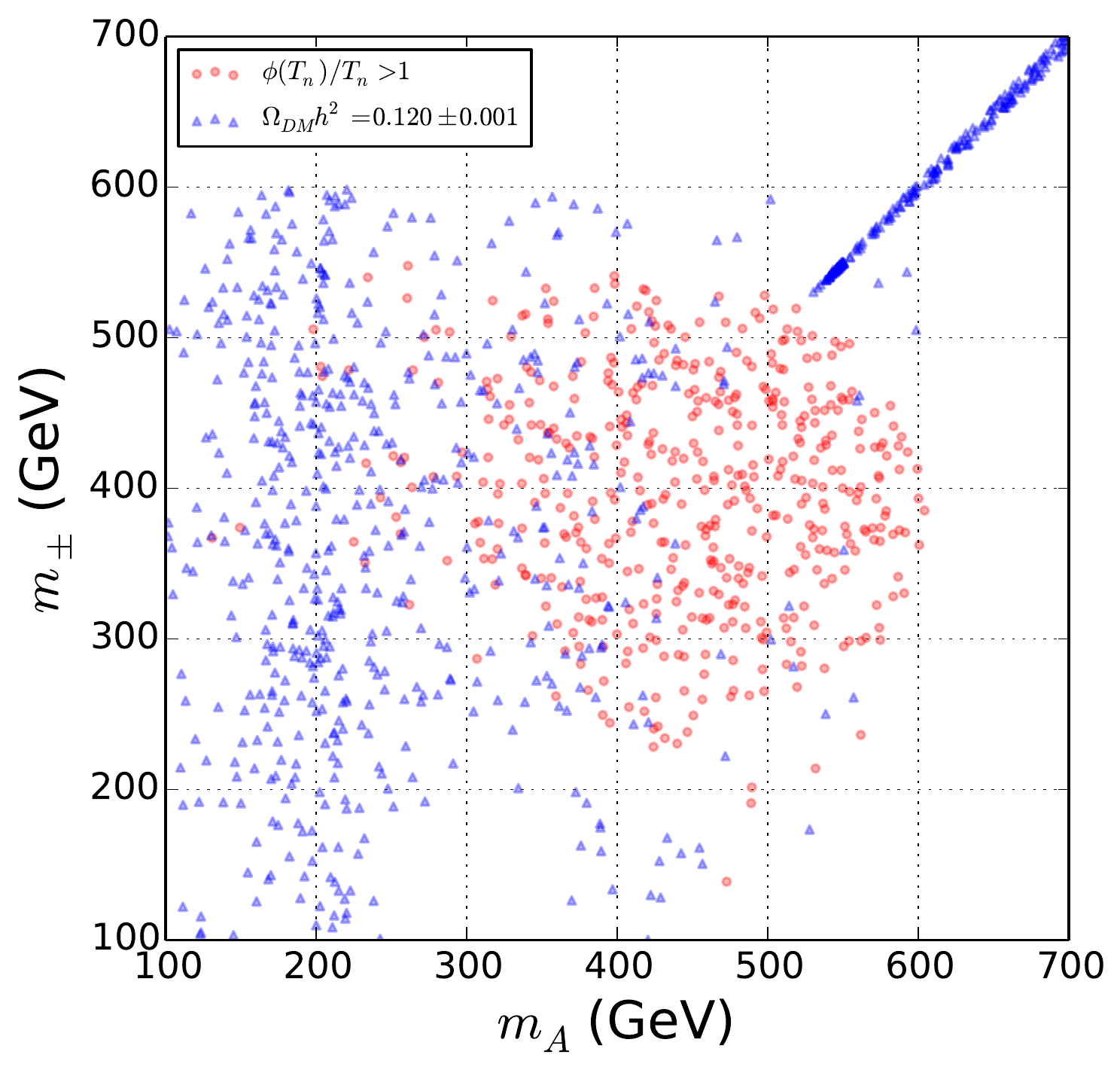}
\caption{Scatter plots on ($m_{\pm},~m_{A}$)-plane satisfying $\phi(T_n) / T_n > 1$ for different scalar DM mass, $m_H$, (upper left) and the strength of the electroweak phase transition $\phi(T_n) / T_n$ (upper right) are shown in the case of scalar DM scenario.
A parameter regime satisfying $\Omega_{\rm DM }h^2 =0.120\pm 0.001$ for different DM mass (lower left) and a combined scatter plot (lower right) for $\phi(T_n) / T_n>1$ (red circles) and $\Omega_{\rm DM }h^2 =0.120\pm 0.001$ (blue triangles) are also shown.
}\label{fig1}
\end{figure}
We show the data points satisfying the conditions $\phi(T_n) / T_n>1$ on ($m_{\pm},~m_A$)-plane in upper panel of Fig.~\ref{fig1}. While DM mass can be low $m_H< 80$ GeV, the charged and neutral pseudo scalar masses are required to be large. In lower left panel of Fig.~\ref{fig1} we show the parameter region on ($m_{\pm},~m_A$)-plane which satisfies $\Omega_{\rm DM }h^2 =0.120\pm 0.001$ with varying DM mass shown as colour code. This clearly shows the two distinct regions of DM (H) mass: $m_{H} <80\;{\rm GeV}$ and $m_{\rm H}>550$ GeV typical of inert doublet dark matter known in the literature. In the lower right panel we also superimpose the points which satisfy the SFOPT criteria, showing overlap with parameter space corresponding to low mass DM. We confirm that there is a parameter regime satisfying conditions $\phi(T_n) / T_n > 1$ and $\Omega_{\rm DM }h^2 =0.120\pm 0.001$, specially for low DM mass $m_{H} <80\;{\rm GeV}$. One should note that the SFOPT requires fine-tuning between $\lambda_1$ and $\lambda_2+2\lambda_3$ to maintain small $\lambda_{H}=\lambda_1+\lambda_2+2\lambda_3$ for a low DM mass regime, $m_{H} <80{\rm GeV}$, as originally pointed out by authors of Ref.~\cite{Borah:2012pu}. While the DM relic satisfying points in low mass regime remain scattered, there is a linear correlation in high mass regime beyond 550 GeV, as can be seen from bottom panel plots of Fig.~\ref{fig1}. This arises because of the fact that, in order to satisfy correct DM relic in high mass regime $m_{\rm H}>550$GeV, the mass splitting between inert doublet components is required to be small. We also find that SFOPT criteria requires the bare mass parameter of inert doublet to be small $0< m_1/{\rm GeV} < 50$. Therefore, in the large DM mass regime, $m_{\rm H}>550$GeV, we need larger values of $m_1$ making the phase transition strength weaker, and thus, it is difficult to realise $\phi(T_n) / T_n >1$ with imposing perturbative conditions $|\lambda_{i}|<4\pi$ ($i=1,2,3$). Thus, the low mass DM region $m_{H} <80{\rm GeV}$ is the preferred region from SFOPT and DM relic point of view in the scalar DM scenario of this model.

\begin{figure}[tbp]
\includegraphics[width=0.45\textwidth]{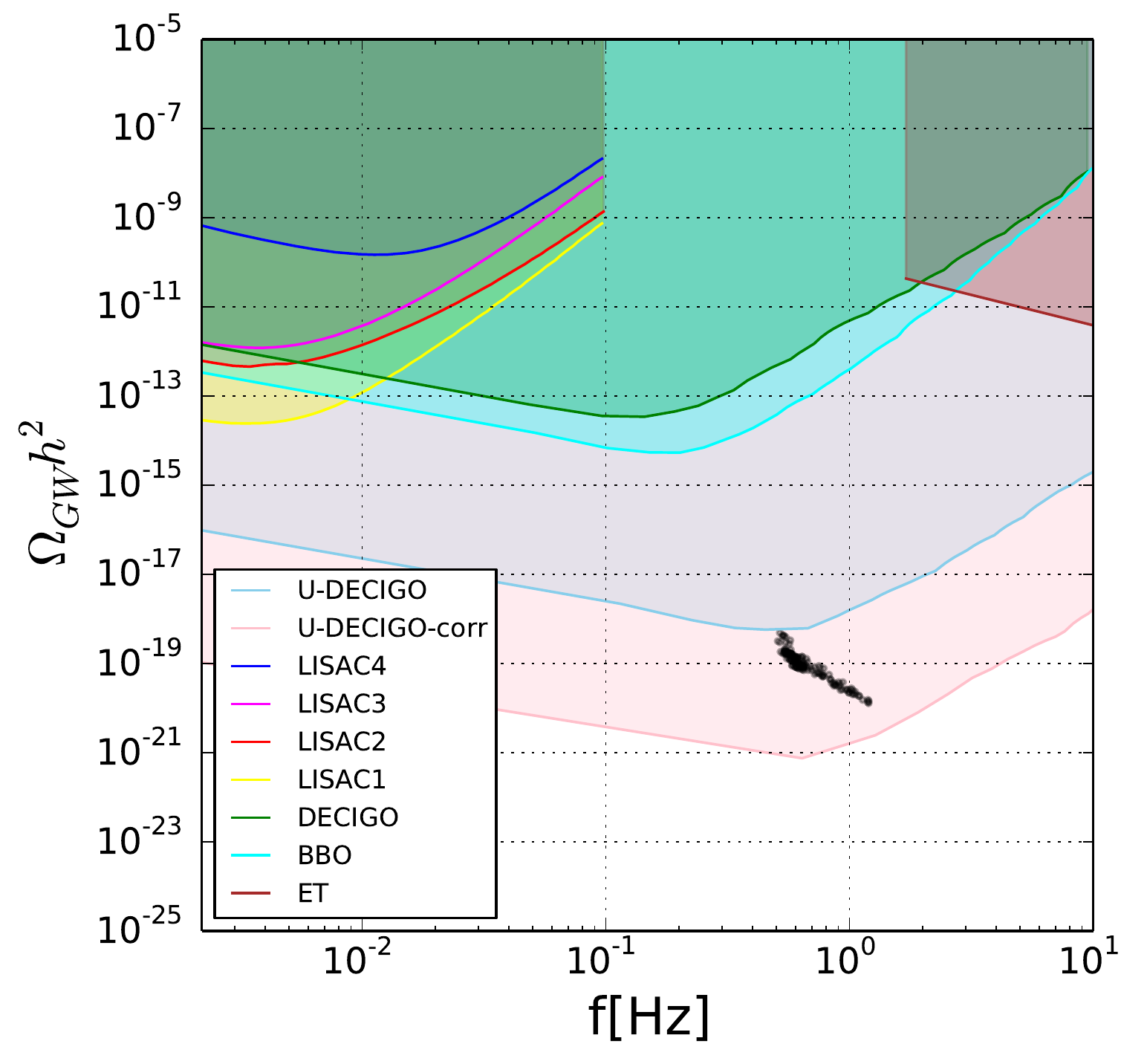}
\includegraphics[width=0.45\textwidth]{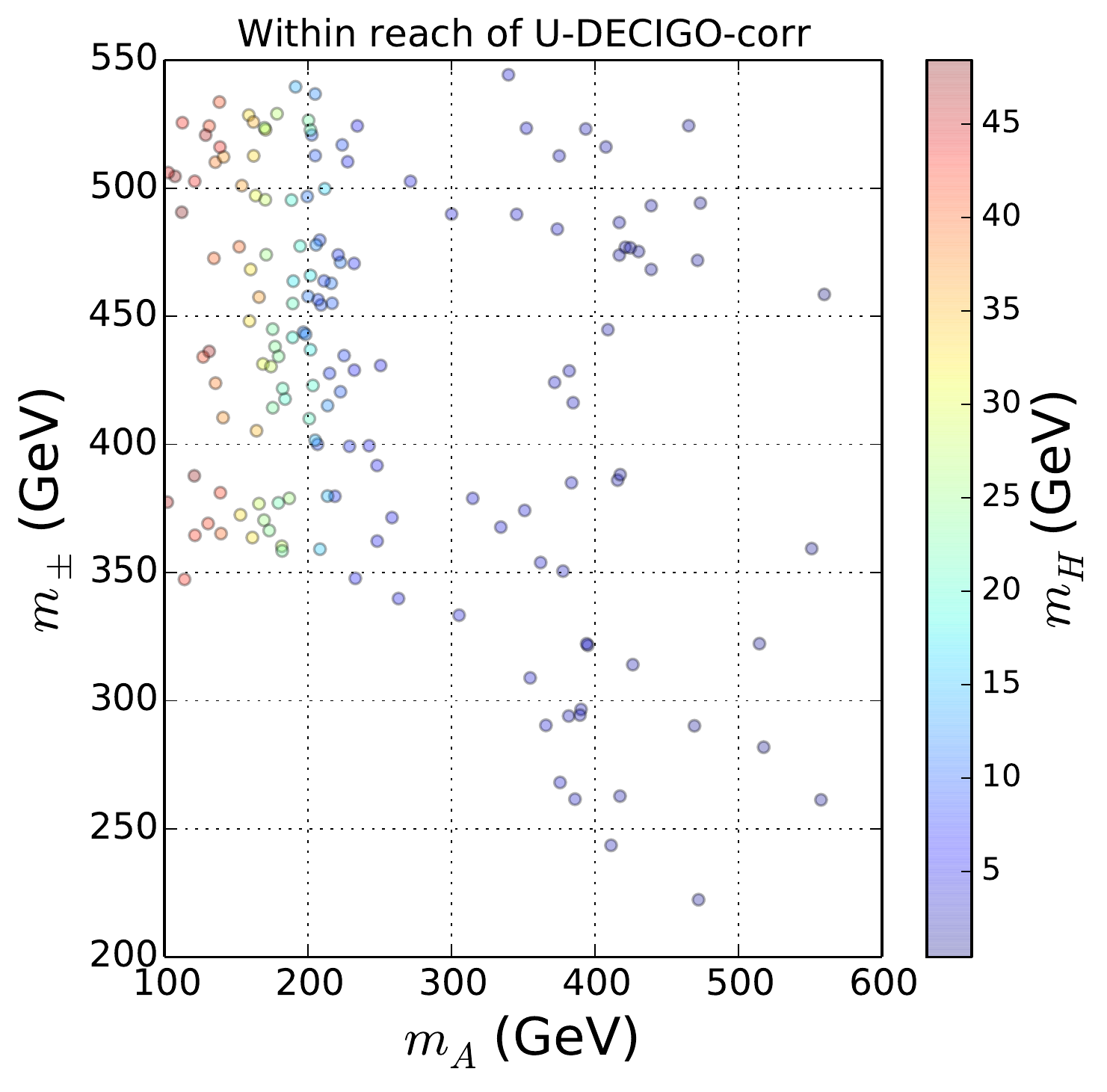}
\caption{
Left panel: peak points of GW signals in our model along with sensitivity curves of U-DECIGO, U-DECIGO-corr, LISAC1$\sim$C4~\cite{Caprini:2015zlo,Klein:2015hvg}, DECIGO, BBO~\cite{Yagi:2011wg} and Einstein Telescope (ET)~\cite{Punturo:2010zz,Hild:2010id}.
Scatter plots on $(m_\pm, m_A)$-plane within reach of U-DECIGO with a correlation analysis (U-DECIGO-corr)~\cite{Kudoh:2005as} for different $m_H$ (right panel) are shown in the case of scalar DM. All data points satisfy correct DM relic density, $\Omega_{\rm DM}h^2 =0.120\pm 0.001$.}
\label{fig2}
\end{figure}
\begin{figure}[tbp]
\includegraphics[width=0.6\textwidth]{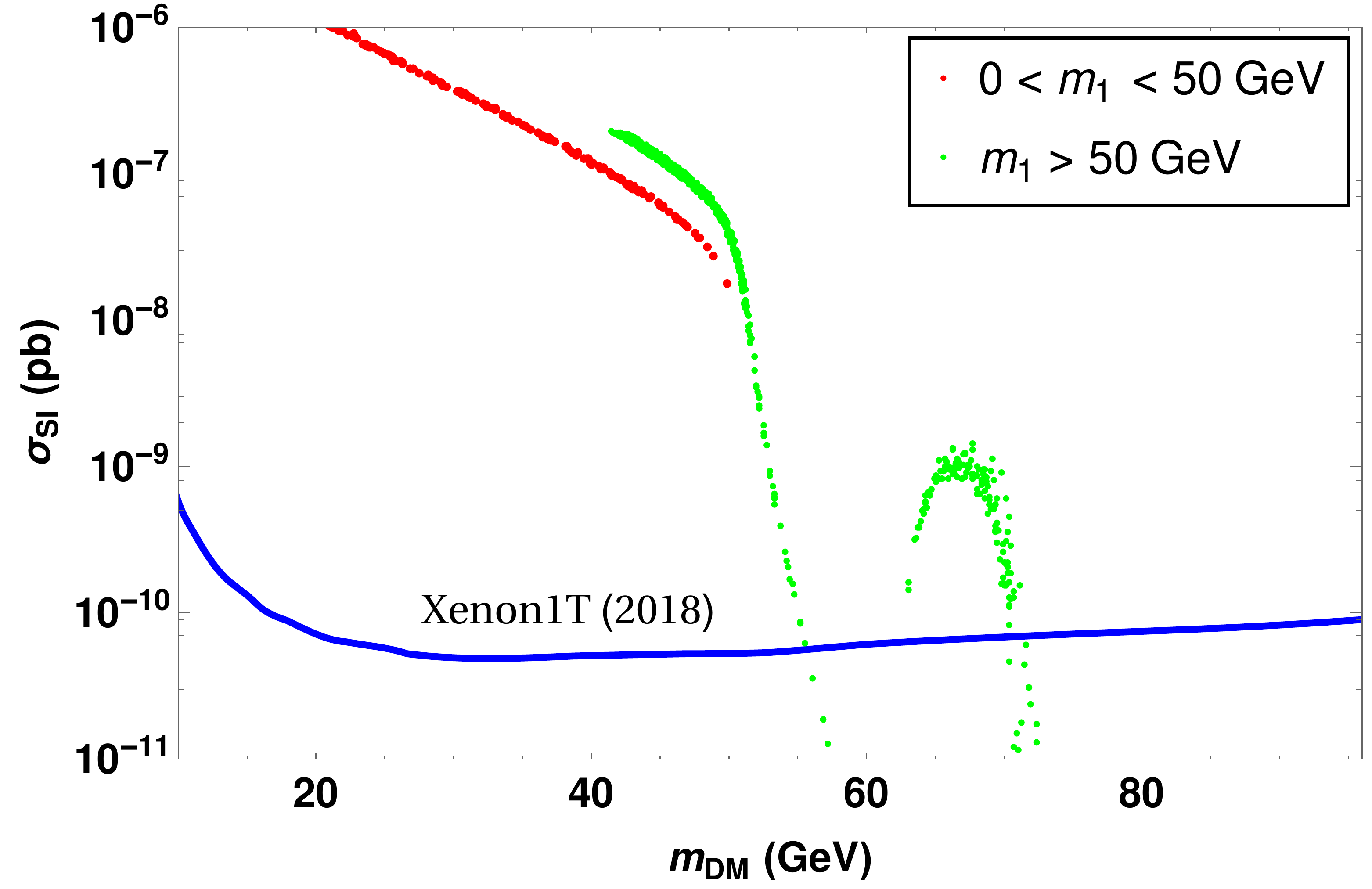}
\caption{Spin independent DM-nucleon cross section for scalar DM. The red coloured points satisfy the SFOPT criteria.}
\label{fig2a}
\end{figure}

In left panel of Fig.~\ref{fig2}, we show total GW signals with sensitivity curves of several future planned space based experiments like U-DECIGO, U-DECIGO-corr, LISA (C1-C4)~\cite{Caprini:2015zlo,Klein:2015hvg}, DECIGO, BBO~\cite{Yagi:2011wg} and Einstein Telescope (ET)~\cite{Punturo:2010zz,Hild:2010id} along with the black dotted region corresponding to our model in scalar DM scenario. Clearly, the strength of the GW signal in our model is within the reach of only U-DECIGO-corr, a future space-based GW antenna \cite{Kudoh:2005as, Sato:2017dkf}. The corresponding region of model parameter space is shown in 
$(m_\pm,m_A)$-plane by taking points within the sensitivity of U-DECIGO-corr (right panel of Fig.~\ref{fig2}). In this figure, the DM relic criteria $\Omega_{\rm DM}h^2 = 0.120\pm 0.001$ is  imposed for all the points.
We find that the total fraction of points in the parameter scan within the reach of U-DECIGO-corr is $39\%$.
However, in this parameter regime, there is a fine-tuning between $\lambda_1$ and $\lambda_2+2\lambda_3$ in order to keep the scalar DM in low mass regime, as noted earlier. We do not investigate such fine-tuned region further in our study. Therefore, in the case of the scalar DM scenario, we conclude that it is difficult to produce detectable GW signals by U-DECIGO-corr while simultaneously giving rise to observed scalar DM relic without fine-tuning. As noted earlier, this low mass DM region is also tightly constrained by direct search and collider experiments. We in fact projected these data points to the direct detection cross section versus DM mass plane and find that they are already disfavoured by Xenon1T data, as seen from figure \ref{fig2a}. While low mass DM is not completely ruled out yet by Xenon1T, the region satisfying SFOPT criteria is ruled out. As can be seen from the green coloured points in this figure, the region around the Higgs resonance $m_{\rm DM}=m_H = m_h/2$ is still allowed. However these green coloured points do not give rise to the required SFOPT. This happens due to the requirement of small $m_1 \in (0, 50) {\rm GeV}$ for SFOPT which further requires larger $\lambda_H$ to keep DM mass near the Higgs resonance area. However, same $\lambda_H$ also leads to spin-independent DM-nucleon scattering giving rise to tension with Xenon1T bounds. As we will see in the next subsection, the fine-tuning associated with scalar DM will be alleviated and stringent direct detection constraints will disappear in the case of the fermion DM scenario.

\subsection{Fermion dark matter}

In this subsection, we show the results in the case of the fermion DM scenario.
Here we do not impose conditions $\lambda_3<0$ and $\lambda_2 +2\lambda_3 <0 $ because as long as lightest singlet fermion $N_1$ is the lightest $Z_2$ odd particle, there is no restriction on the mass ordering among components of inert scalar doublet.

\begin{figure}[tbp]
\includegraphics[width=0.45\textwidth]{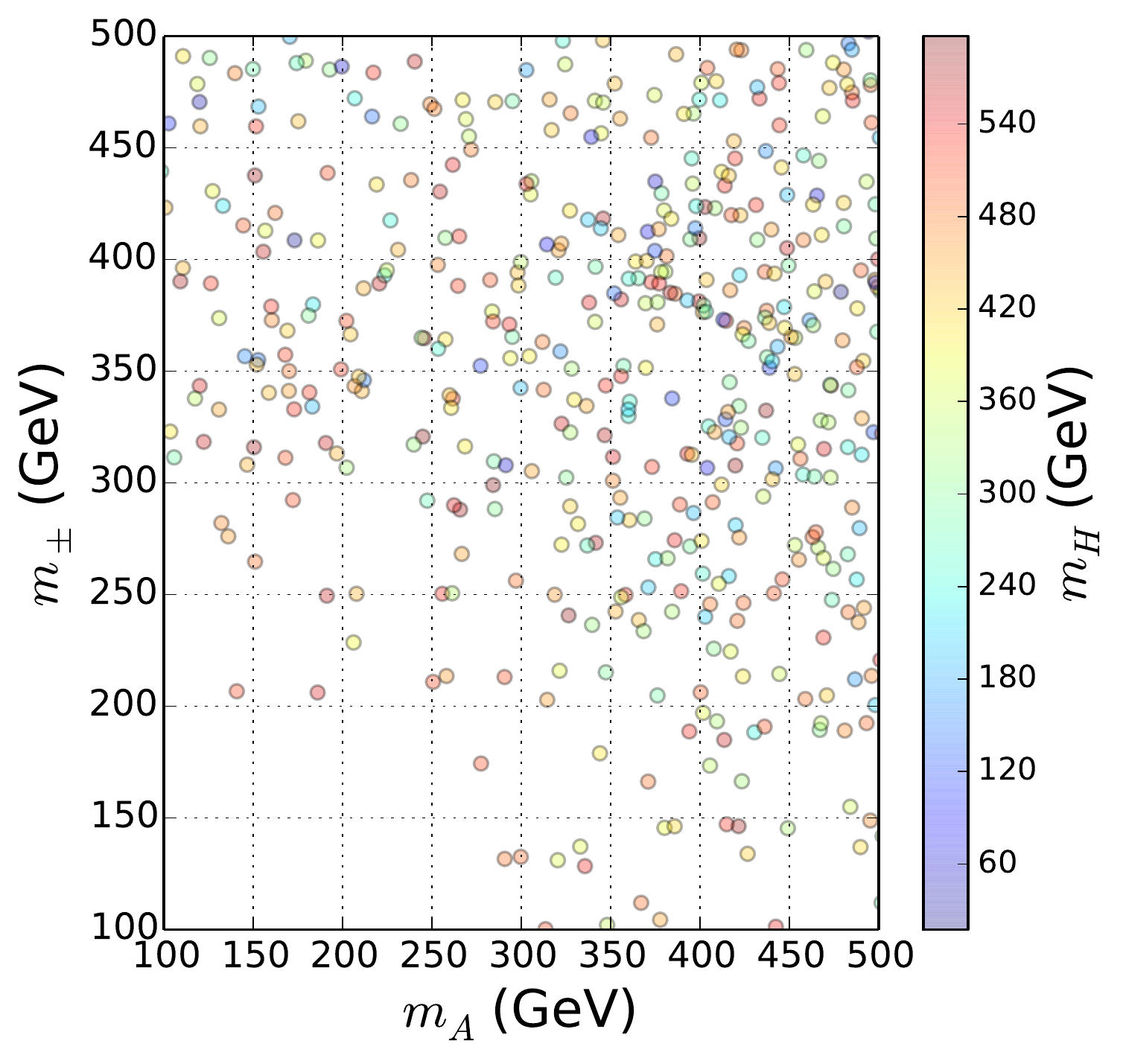}
\includegraphics[width=0.45\textwidth]{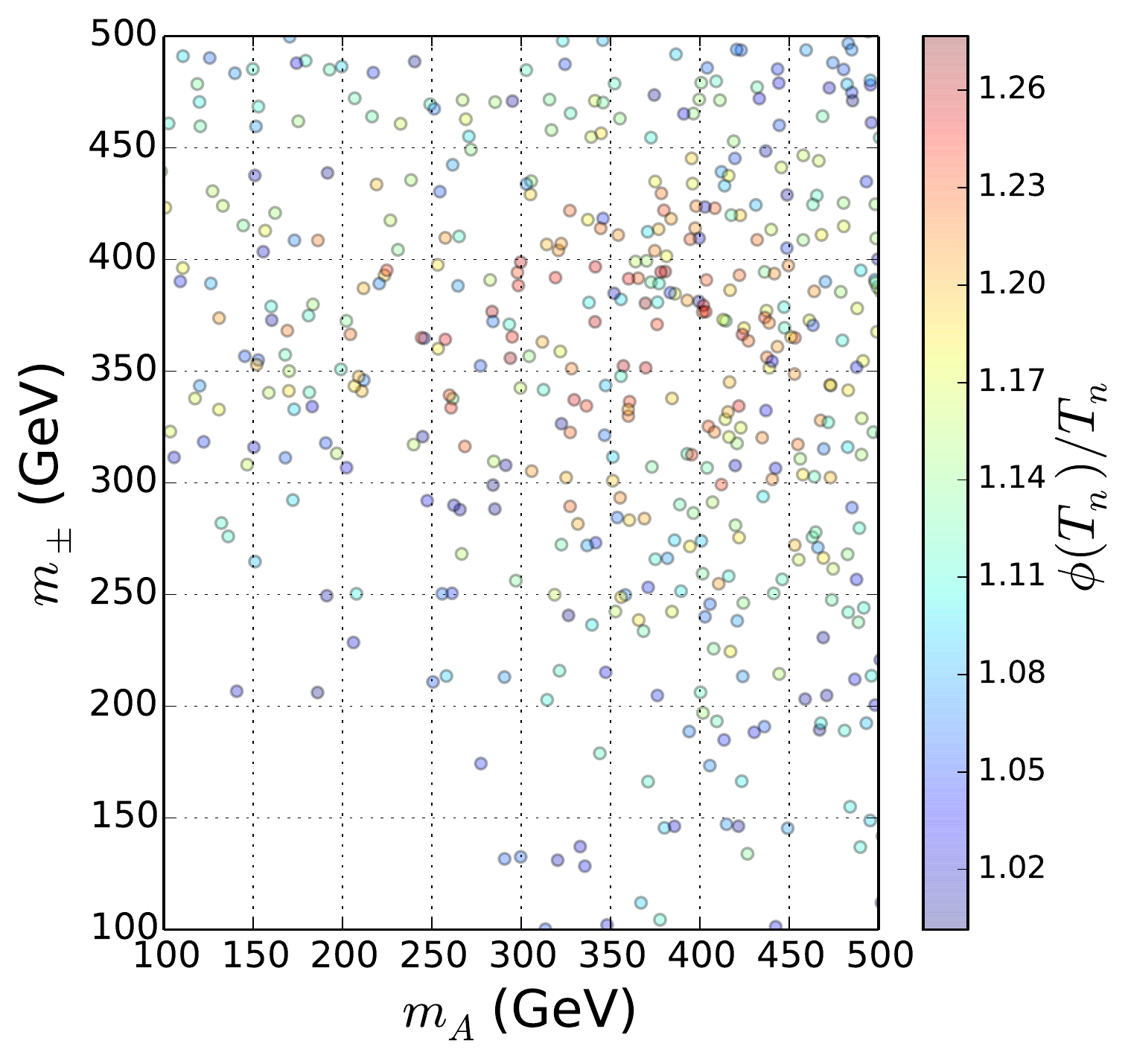}
\includegraphics[width=0.45\textwidth]{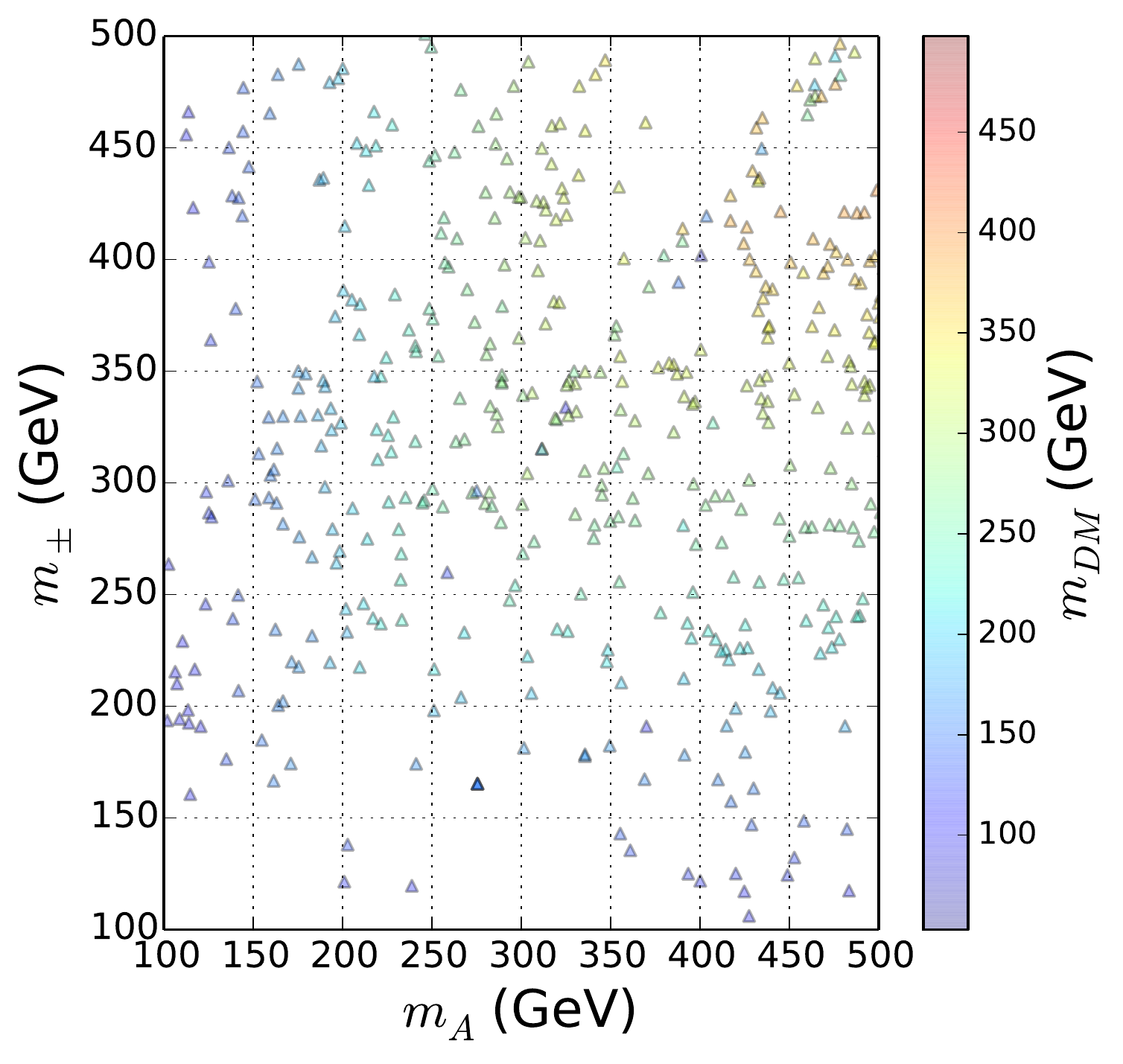}
\includegraphics[width=0.45\textwidth]{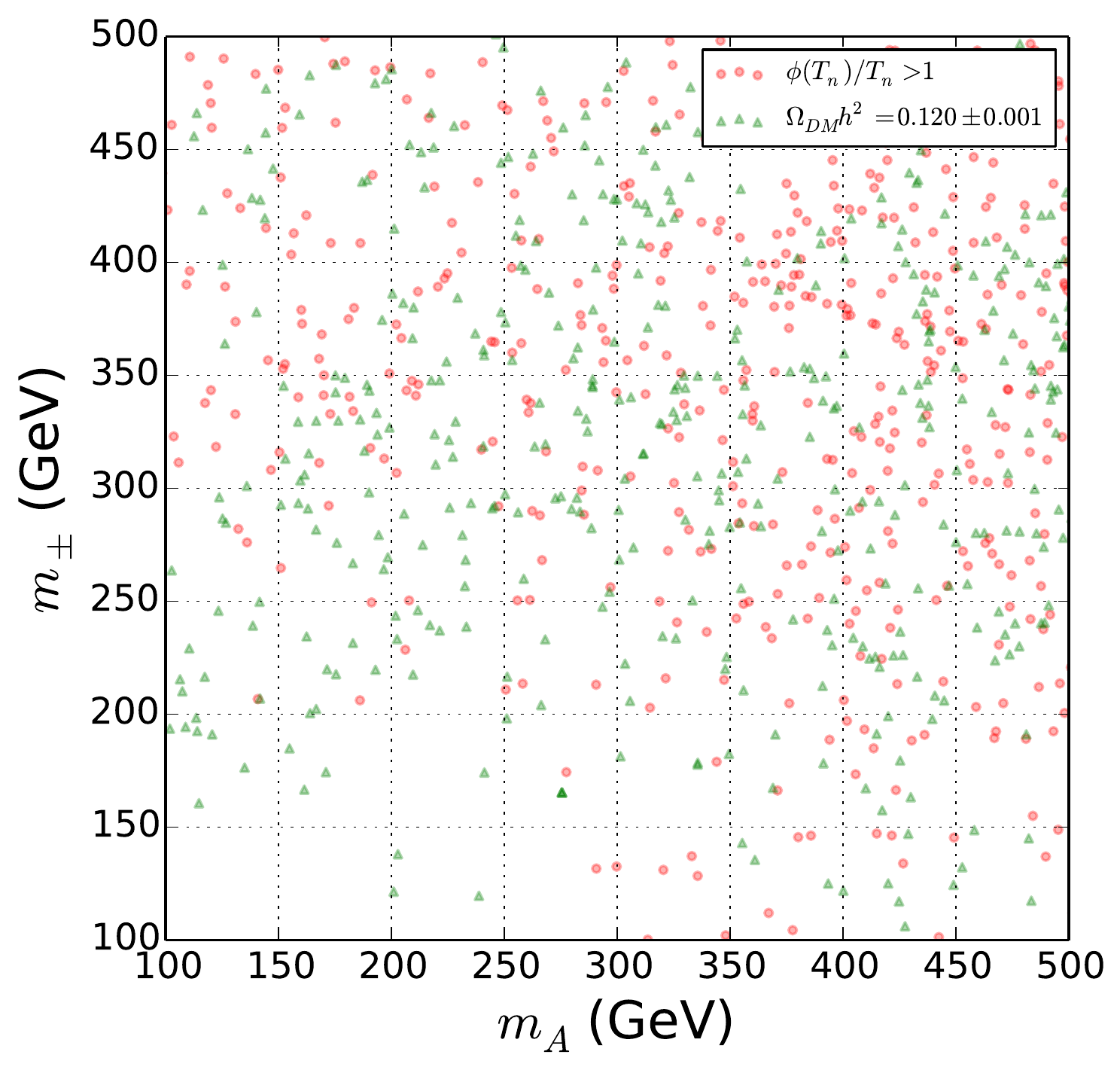}
\caption{Scatter plots on ($m_{\pm},~m_{A}$)-plane satisfying $\phi(T_n) / T_n > 1$ for different $m_H$, (upper left) and the strength of the electroweak phase transition $\phi(T_n) / T_n$ (upper right) are shown in the case of fermion DM scenario.
A parameter regime satisfying $\Omega_{\rm DM }h^2 =0.120\pm 0.001$ for different $m_H$ (lower left) and a combined scatter plot (lower right) for $\phi(T_n) / T_n>1$ (red circles) and $\Omega_{\rm DM }h^2 =0.120\pm 0.001$ (green triangles) are also shown.
}\label{fig3}
\end{figure}
We show the parameter regime satisfying the conditions $\phi(T_n) / T_n>1$ in ($m_{\pm},~m_A$)-plane in upper panel of Fig.~\ref{fig3} in the case of the fermion DM scenario.
As one can see from upper panels of Fig.~\ref{fig3}, we additionally have a parameter regime satisfying $\phi(T_n) /T_n >1$ for small values of $m_{\pm}$ and $m_A $ compared to what we found in the scalar DM scenario.
This fact can be understood as following way. Since $\phi(T_n) / T_n >1$ can be realised for smaller $\lambda_1$ (corresponding to smaller $m_\pm$) by making $\lambda_H$ larger (corresponding to larger $m_{H}$) with fixed $m_1$, a lower $m_{\pm}$ region can lead to $\phi(T_n) / T_n >1$ in comparison to the scalar DM scenario. This is a crucial difference from similar SFOPT analysis in IDM where the mass ordering within inert doublet components is always restricted to having one of the neutral components as the lightest. While sub-dominant scalar DM leads to more allowed region satisfying SFOPT as shown by \cite{Cline:2013bln}, relaxing the mass ordering leads to even newer allowed region of parameter space. This has been made possible in scotogenic model where fermion DM remains a viable possibility. While in the upper panel plots of Fig.~\ref{fig3}, the DM relic constraint is not implemented, we do so in the lower panel plots of the same figure. The lower left panel shows the parameter space in ($m_{\pm},~m_A$)-plane which satisfied the correct fermion DM relic while showing the fermion DM mass in colour code. The right lower panel plot shows the points satisfying SFOPT and DM relic criteria on ($m_{\pm},~m_A$)-plane clearly depicting overlap where both are satisfied.
The points satisfying the fermion DM relic correspond to small $\lambda_3$ so that the Yukawa couplings (in $Y_{i1} \, \bar{L}_i \tilde{S} N_1$) are sizeable enough to enhance fermion DM annihilation and coannihilation channels. This correspondence between small $\lambda_3$ and large Yukawa arises through Casas-Ibarra parametrisation discussed earlier. We consider lightest neutrino mass $0.1$ eV in order to enhance the Yukawa couplings. Small $\lambda_3$ gives rise to almost degenerate $H, A$ in this scenario.
\begin{figure}[tbp]
\includegraphics[width=0.45\textwidth]{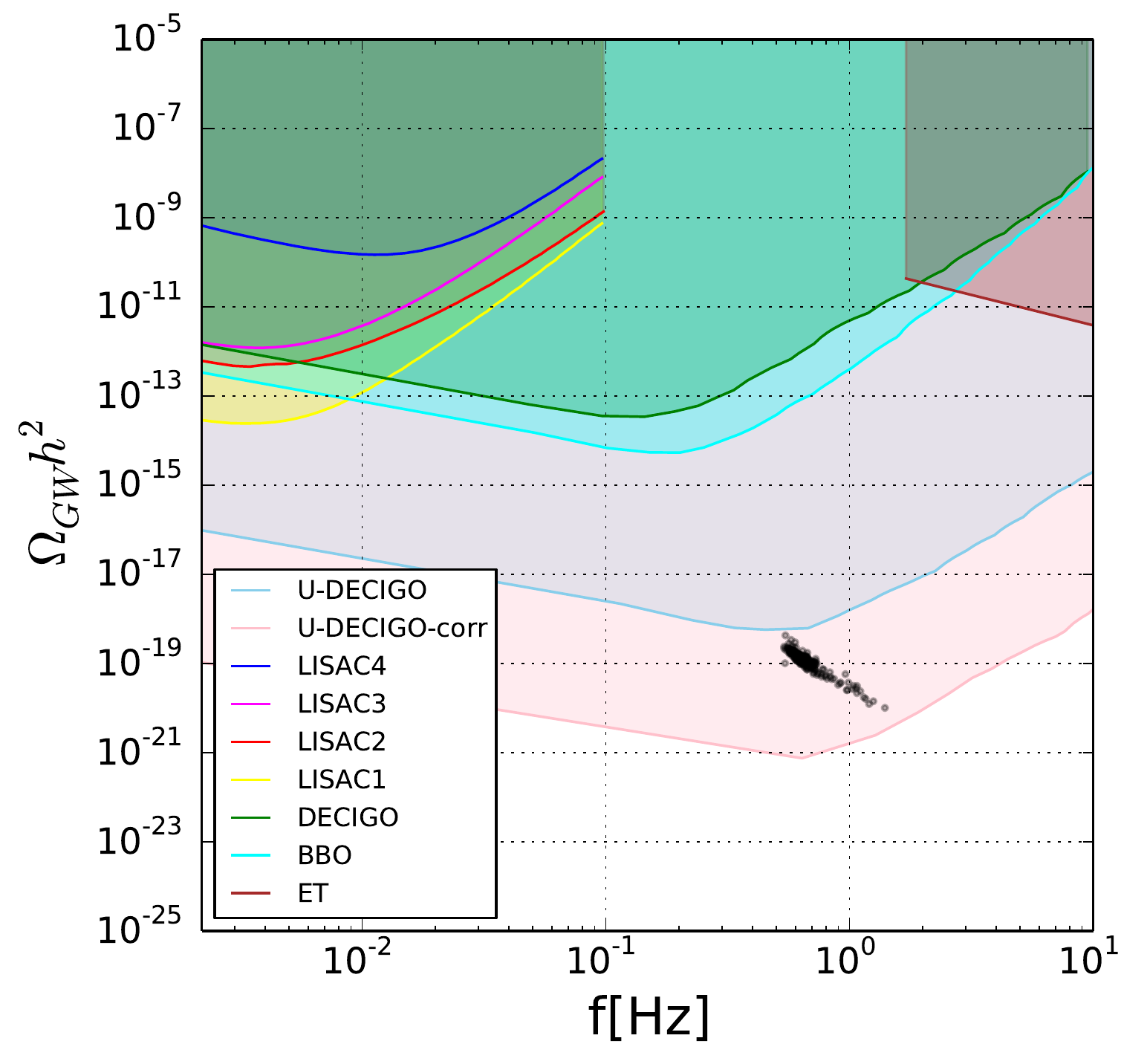}
\includegraphics[width=0.45\textwidth]{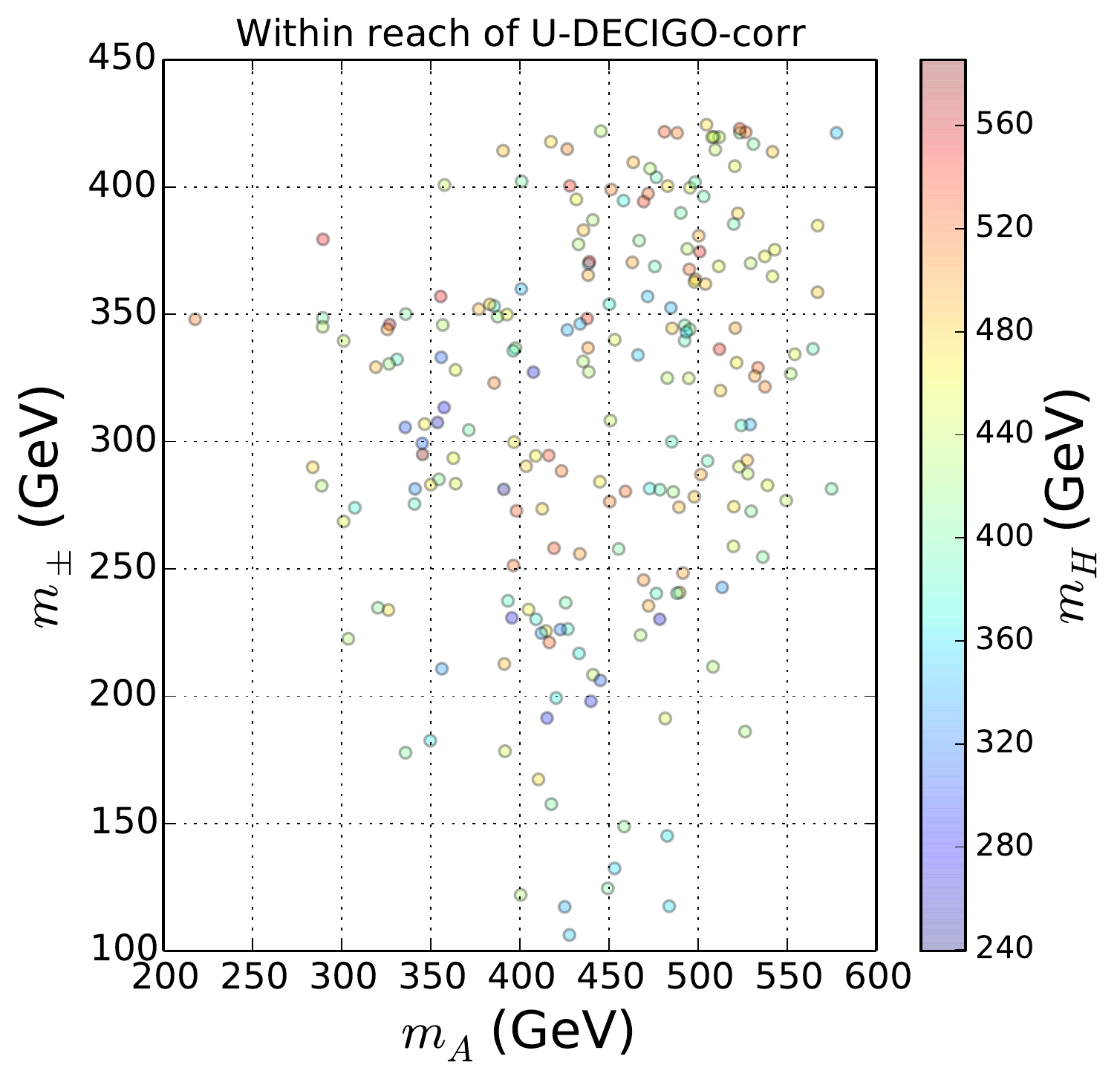}
\caption{Left panel: peak points of GW signals along with sensitivity curves of U-DECIGO, U-DECIGO-corr, LISAC1$\sim$C4, DECIGO, BBO and ET. Scatter plots on $(m_\pm, m_A)$-plane within reach of U-DECIGO-corr for different $m_H$ (right panel) is shown in the case of fermion DM scenario. All data points satisfy correct DM relic density, $\Omega_{\rm DM}h^2 =0.120\pm 0.001$.}
\label{fig4}
\end{figure}

We also show peak amplitudes of total GW signals in our model with fermion DM scenario along with sensitivity curves of different planned future experiments in left panel of Fig.~\ref{fig4}. We then show scatter plots in $(m_\pm,m_A)$-plane by taking points within reach of U-DECIGO-corr and satisfying $\Omega_{\rm DM }h^2 =0.120\pm 0.001$ in right panel of Fig.~\ref{fig4}, respectively. Since there is no restriction on the scalar mass ordering within inert scalar doublet in the case of the fermion DM scenario, GW signals can be easily enhanced by making $\lambda_H$ large compared to the scalar DM scenario. Indeed, we find that the total fraction of points in the parameter scan within the reach of U-DECIGO-corr is $58\%$, which is larger than the one in the case of scalar DM scenario. Unlike in scalar DM scenario, such large $\lambda_H$ does not contribute to direct detection of fermion DM. In fact, due to gauge singlet and leptophilic nature of fermion DM there is no tree level direct detection cross section, keeping it safe from Xenon1T bounds. It should also be emphasised that we have no fine-tuning between $\lambda_1,~\lambda_2$ and $\lambda_3$ in this scenario, and thus, detectable GW signals at U-DECIGO-corr are naturally produced. However, such leptophilic fermion DM scenario can be tightly constrained by experimental bounds on charged lepton flavour violation as we discuss below.

\subsection{Charged Lepton Flavour Violation}
Charged lepton flavour violation (CLFV) arises in the SM at one loop level and remains suppressed by the smallness of neutrino masses, much beyond the current and near future experimental sensitivities. Therefore, any experimental observation of such processes is definitely a sign of BSM physics, like the one we are studying here. In the present model, this becomes inevitable due to the couplings of new $Z_2$ odd particles to the SM lepton doublets. The same fields that take part in the one-loop generation of light neutrino mass, can also mediate charged lepton flavour violating processes like $\mu \rightarrow e \gamma, \mu \rightarrow 3e$ and $\mu \rightarrow e$ (Ti) conversion. These rare processes have strong current limit as well as good future sensitivity \cite{Toma:2013zsa}. The present bounds are: ${\rm BR}(\mu \rightarrow e \gamma) < 4.2 \times 10^{-13}$ \cite{TheMEG:2016wtm},  ${\rm BR}(\mu \rightarrow 3e) < 1.0 \times 10^{-12}$ \cite{Bellgardt:1987du}, ${\rm CR} (\mu, \rm Ti \rightarrow e, \rm Ti) < 4.3 \times 10^{-12}$ \cite{Dohmen:1993mp}. While the future sensitivity of the first two processes are around one order of magnitude lower than the present branching ratios, the $\mu$ to $e$ conversion (Ti) sensitivity is supposed to increase by six order of magnitudes \cite{Toma:2013zsa} making it a highly promising test to confirm or rule out different TeV scale BSM scenarios.

\begin{figure}
\centering
    \includegraphics[width=0.48\textwidth]{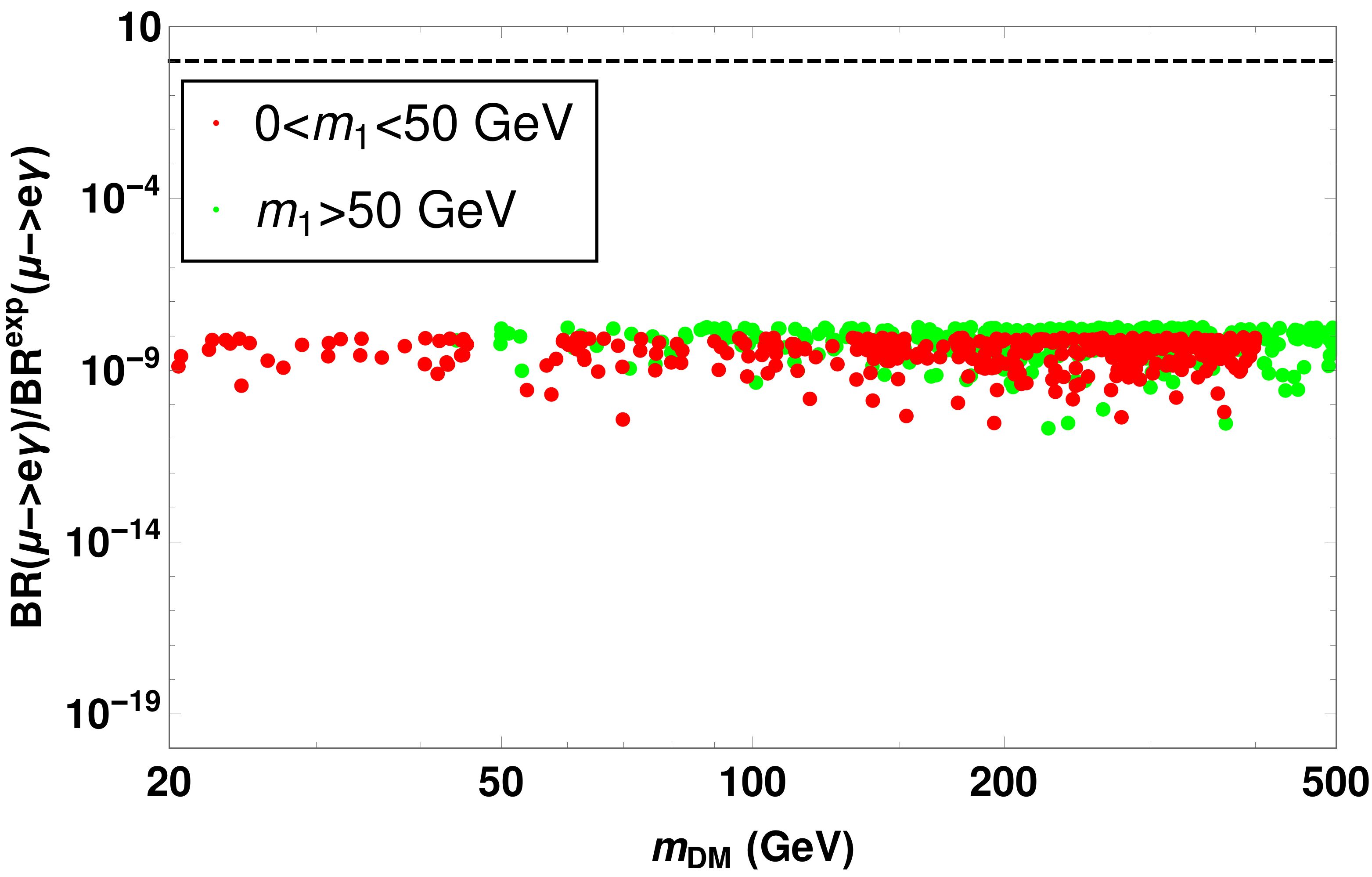} 
    \includegraphics[width=0.48\textwidth]{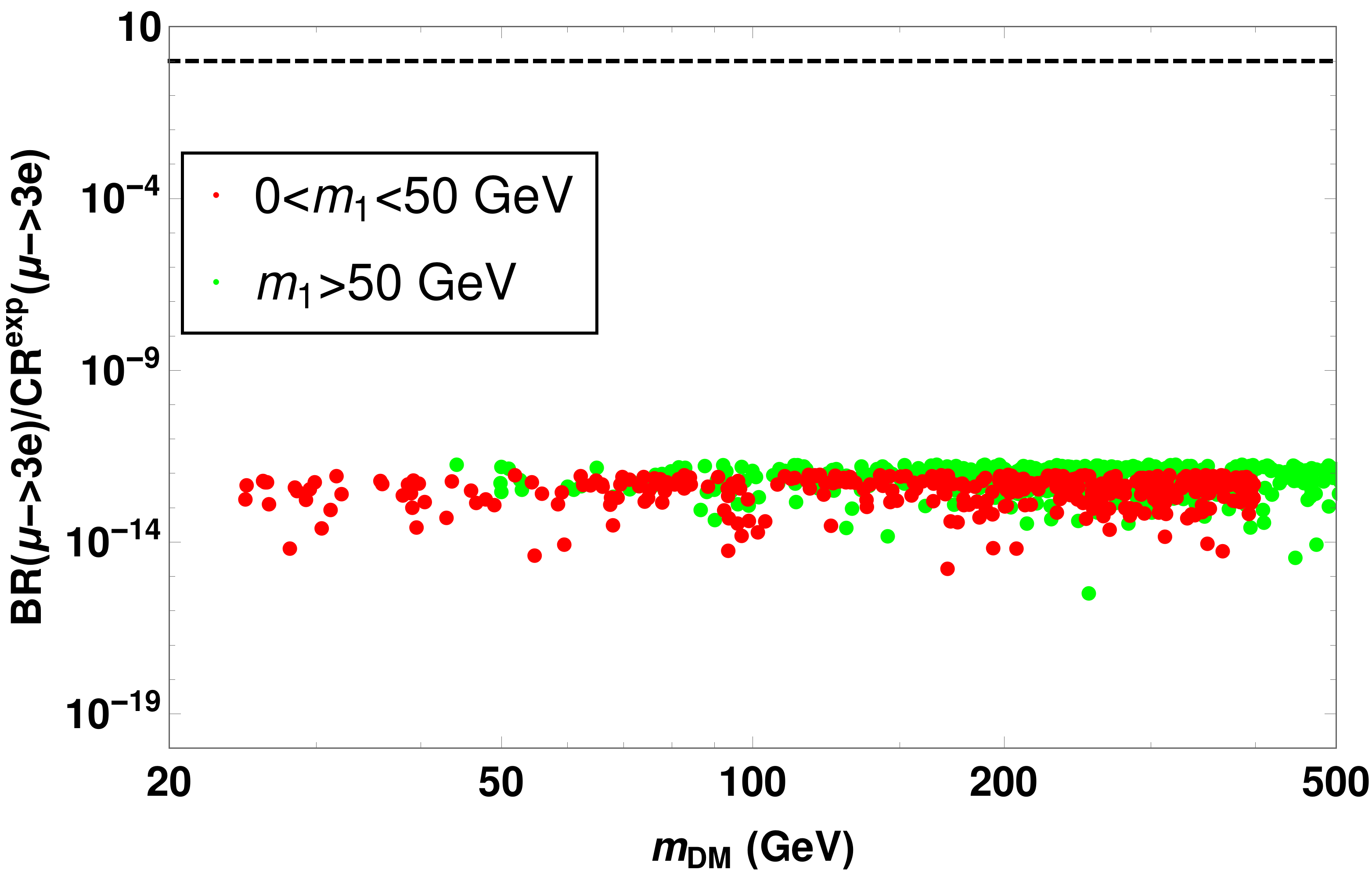} \\
     \includegraphics[width=0.48\textwidth]{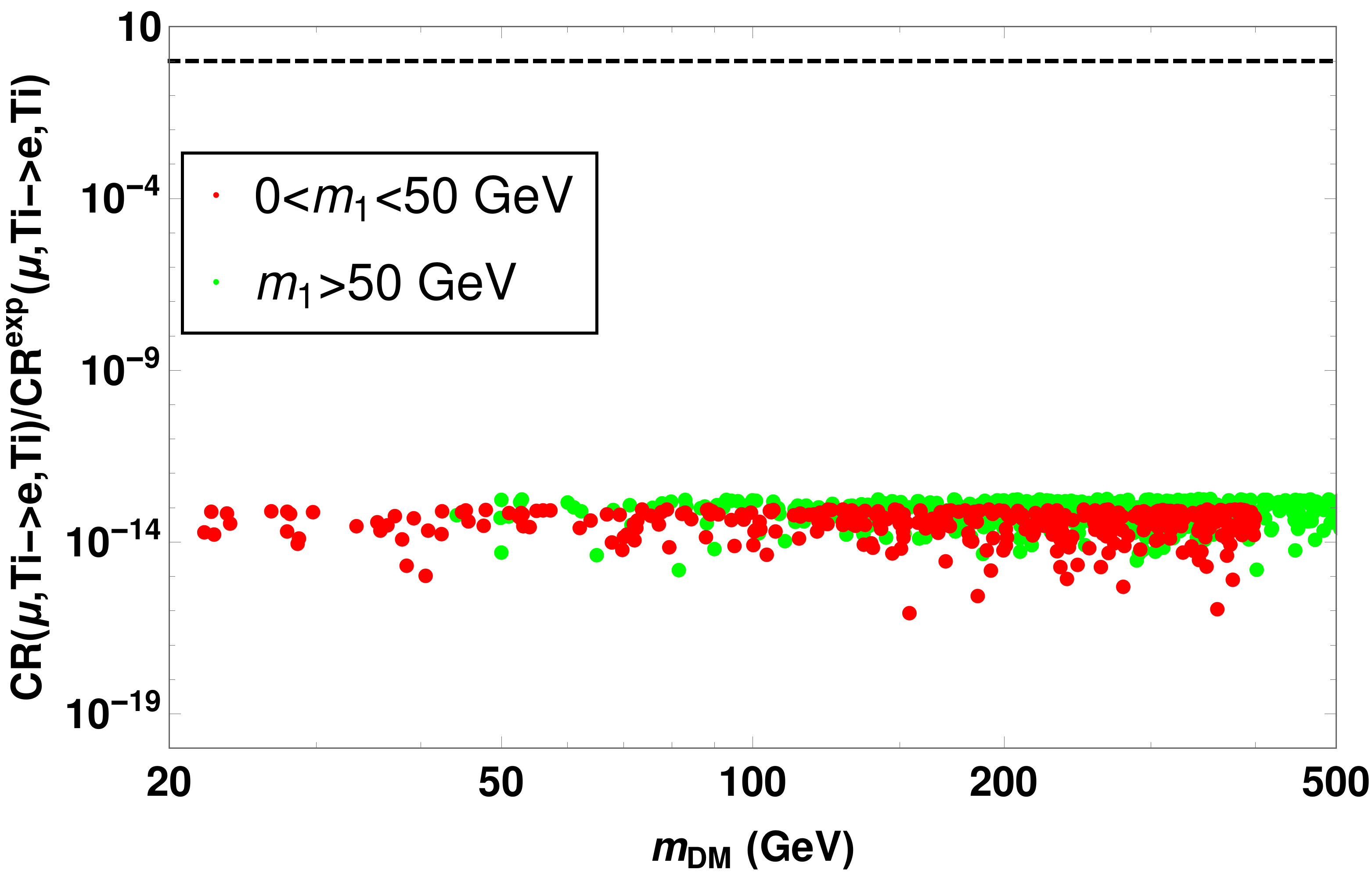} 

 \caption{Predictions for different CLFV processes for fermion DM scenario. All points satisfy the DM relic criteria while the red coloured points satisfy the SFOPT criteria.}
 \label{lfvplot}
\end{figure}

Since the couplings, masses involved in this process are the same as the ones that generate light neutrino masses and play a role in DM relic abundance, we can no longer choose them arbitrarily. The branching fraction for $\mu \rightarrow e \gamma$ that follows from this one-loop contribution can be written as~\cite{Vicente:2014wga},
\begin{equation}
{\rm Br}(\mu \rightarrow e \gamma) = \frac{3(4 \pi)^{3} \alpha_{\rm em}}{4 G_{F}^{2}}|A_{D}|^{2} {\rm Br}(\mu \rightarrow e \nu_{\mu} \bar{\nu_{e}}).
\label{br_meg}
\end{equation}
Where $\alpha_{\rm em}$ is the electromagnetic fine structure constant, $e$ is the electromagnetic coupling and $G_{F}$ is the Fermi constant. $A_{D}$ is the dipole form factor given by
\begin{equation}
A_{D} = \sum_{i=1}^{3} \frac{Y_{ie}^{*}Y_{i\mu}}{2(4\pi)^{2}} \frac{1}{m^2_{\pm}}\left(\frac{1-6 \xi_{i}+3 \xi_{i}^{2}+2 \xi_{i}^{3}-6 \xi_{i}^{2} log\xi_{i}}{6(1-\xi_{i})^{4}}\right).
\end{equation}
Here the parameter $\xi_{i}$'s are defined as $\xi_{i} \equiv M_{N_{i}}^{2}/m^2_{\pm}$. The MEG experiment provides the most stringent upper limit on the branching ratio ${\rm Br}(\mu \rightarrow e \gamma) < 4.2 \times 10^{-13}$ \cite{TheMEG:2016wtm}. For analytical expressions of other flavour violating processes, please refer to \cite{Vicente:2014wga}. We have used the \texttt{SPheno 3.1} interface \cite{Porod:2011nf} in order to implement the flavour constraints into the model. For fermion DM model, the predictions for CLFV processes are shown in figure \ref{lfvplot}. As can be seen from this figure, all the predicted points lie way below the current experimental bounds. This is due to the fact that fermion singlet DM relic is mainly governed by its coannihilation with inert doublet components and hence even relatively smaller Yukawa couplings can give rise to the correct relic. This was also noted in earlier works \cite{Vicente:2014wga}.

\section{Conclusion}
\label{sec:conclude}
We have studied the possibility of generating GWs from a strong first-order EWPT in minimal scotogenic model which can be probed at future space based experiments like U-DECIGO. While the scalar content of the model is same as that in IDM  where the possibility of strong first-order EWPT along with scalar DM has been studied in several earlier works, in the present model we find newer region of parameter space due to the possibility of fermion DM. We also improve earlier studies by appropriate consideration of resummation effects in finite temperature effective potential. Our results in the scalar DM scenario is partially in agreement with earlier works where SFOPT criteria favours a low mass scalar DM which however, remains disfavoured from stringent direct detection bounds from Xenon1T 2018. The parameter space favouring SFOPT can however be enhanced for sub-dominant scalar DM, in agreement with earlier works. More interesting results arise in the fermion DM scenario where the SFOPT favoured parameter space is enlarged primarily due to the fact that mass ordering within the inert scalar doublet components is relaxed in this scenario. This is contrast with the scalar DM scenario where one of the neutral components was restricted to be the lightest $Z_2$ odd particle. Also, unlike scalar DM which can satisfy relic in two well defined regions: low mass regime $m_H <80$ GeV and high mass regime $m_H >550$ GeV, fermion DM (of thermal WIMP type) relic can be realised for almost any mass, as long as the mass difference between fermion DM and next to lightest $Z_2$ odd particle is kept small (to enhance coannihilation) and corresponding Yukawa coupling is kept sizeable in agreement with light neutrino masses via Casas-Ibarra parametrisation. The fermion DM scenario therefore, can also be tightly related to light neutrino masses as the same Yukawa couplings go into the light neutrino mass generation at one loop level. While leptophilic fermion DM in this model can also give rise to sizeable charged lepton flavour violation, we find that for the parameter space satisfying SFOPT, DM relic and other relevant bounds, the contribution to charged rare days remains well below current limits.The possibility of enlarged parameter space in the fermion DM scenario specially having light charged scalars can give rise to interesting signatures at colliders. Such light charged scalars can be produced significantly and can leave interesting signatures like displaced vertex, as discussed recently in \cite{Borah:2018smz}. Connection of such strong first-order EWPT to baryogenesis is another interesting possibility which we leave for future studies. While baryogenesis through leptogenesis is already a viable possibility in the minimal scotogenic model \cite{Hugle:2018qbw, Borah:2018rca,Huang:2018vcr, Baumholzer:2018sfb, Borah:2018uci, Mahanta:2019gfe, Mahanta:2019sfo} as mentioned earlier, a GW based probe of this model can indirectly probe the parameter space relevant for successful leptogenesis as well. Similar ways of probing seesaw and leptogenesis have also been proposed recently \cite{Dror:2019syi, Blasi:2020wpy}. However, possible extensions of the scotogenic model in order to implement electroweak baryogenesis may also bring the GW signal within reach of other upcoming experiments like LISA. We leave such detailed studies in connection to baryogenesis to future works.

\acknowledgments
DB acknowledges the support from Early Career Research Award from the Science and Engineering Research Board (SERB), Department of Science and Technology (DST), Government of India (reference number: ECR/2017/001873). The work of SKK and AD was supported by the NRF of Korea Grants No. 2017K1A3A7A09016430, and No. 2017R1A2B4006338. The work of KF was supported by JSPS Grants-in-Aid for Research Fellows No. 20J12415. KF was also supported by JSPS and NRF under the Japan-Korea Basic Scientific Cooperation Program and would like to thank participants attending the JSPS and NRF conference for useful comments. DB and KF thank the organisers of {\it TokyoTech and IIT Guwahati Joint Workshop: Condensed Matter and High-Energy Physics} at TokyoTech during November 11-12, 2019 where part of this work was discussed.

%

\providecommand{\href}[2]{#2}\begingroup\raggedright\endgroup

\end{document}